\PassOptionsToPackage{table,xcdraw}{xcolor}

\documentclass[sigconf]{acmart}

\usepackage{hyperref}
\usepackage{nameref}
\usepackage{multirow}
\usepackage{tcolorbox}
\definecolor{light-gray}{gray}{0.85}

\AtBeginDocument{%
  \providecommand\BibTeX{{%
    \normalfont B\kern-0.5em{\scshape i\kern-0.25em b}\kern-0.8em\TeX}}}

\copyrightyear{2024}
\acmYear{2024}
\setcopyright{acmlicensed}
\acmConference[CHI '24]{Proceedings of the CHI Conference on Human Factors in Computing Systems}{May 11--16, 2024}{Honolulu, HI, USA}
\acmBooktitle{Proceedings of the CHI Conference on Human Factors in Computing Systems (CHI '24), May 11--16, 2024, Honolulu, HI, USA}
\acmDOI{10.1145/3613904.3642465}
\acmISBN{979-8-4007-0330-0/24/05}

\acmSubmissionID{5099}

\begin{document}


\title{The Adaptive Architectural Layout: How the Control of a Semi-Autonomous Mobile Robotic Partition was Shared to Mediate the Environmental Demands and Resources of an Open-Plan Office}
\renewcommand{\shorttitle}{The Adaptive Architecture Layout}

\author{Binh Vinh Duc Nguyen}
\email{alex.nguyen@kuleuven.be}
\orcid{0000-0001-5026-474X}
\affiliation{
  \institution{Research[x]Design, Department of Architecture, KU Leuven}
  \streetaddress{Kasteelpark Arenberg 1 - box 2431}
  \city{Leuven}
  \country{Belgium}
  \postcode{3001}
}

\author{Andrew Vande Moere}
\email{andrew.vandemoere@kuleuven.be}
\orcid{0000-0002-0085-4941}
\affiliation{
  \institution{Research[x]Design, Department of Architecture, KU Leuven}
  \streetaddress{Kasteelpark Arenberg 1 - box 2431}
  \city{Leuven}
  \country{Belgium}
  \postcode{3001}
}


\begin{abstract}
A typical open-plan office layout is unable to optimally host multiple collocated work activities, personal needs, and situational events, as its space exerts a range of environmental demands  on workers in terms of maintaining their acoustic, visual or privacy comfort. As we hypothesise that these demands could be coped by optimising the environmental resources of the architectural layout, we deployed a mobile robotic partition that autonomously manoeuvres between predetermined locations. During a five-weeks in-the-wild study within a real-world open-plan office, we studied how 13 workers adopted four distinct adaptation strategies when sharing the spatiotemporal control of the robotic partition. Based on their logged and self-reported reasoning, we present six initiation regulating factors that determine the appropriateness of each adaptation strategy. This study thus contributes to how future human-building interaction could autonomously improve the experience, comfort, performance, and even the health and wellbeing of multiple workers that share the same workplace. 
\end{abstract}



\begin{teaserfigure}
  \centering
  \includegraphics[width=0.9\textwidth]{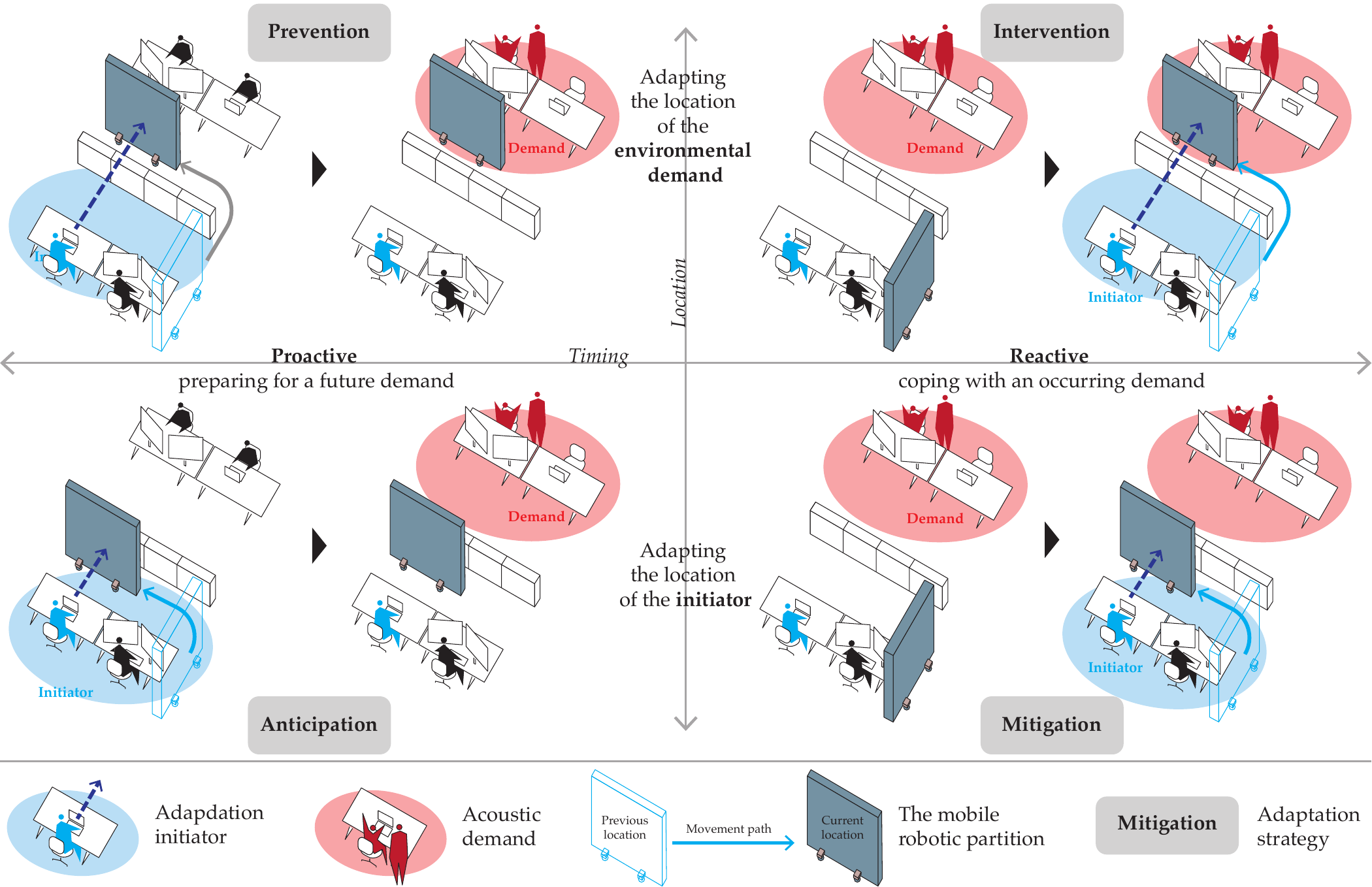}
  \caption{
  The four different spatiotemporal adaptation strategies that were deployed by individual workers to address the environmental demands in their open-plan office. We propose to distribute the strategies across two axes: timing (i.e. before versus during the occurrence of a demand), and location (at the location of the demand versus the worker who initiated the adaptation).
  }
  \Description{A diagrammatic quadrant of four different spatiotemporal adaptation strategies. Each strategy is illustrated through a two-phase sequence depicting how a worker, referred to as the adaptation initiator, initiates an adaptation using a robotic partition to cope with an acoustic demand arising from collaborative activities of two other workers in the same office. Accordingly, the initiator can choose to proactively initiate an adaptation around the their own location to anticipate the potential acoustic demand, or around the location of the workers responsible for the acoustic demand to prevent its potential harm. They may also reactively initiate an adaptation around themselves to mitigate the impact when the acoustic demand is already occurring, or around the location of the workers responsible for the acoustic demand to intervene on their demand-causing activity.}
  \label{fig:teaser}
\end{teaserfigure}


\maketitle

\section{Introduction}
It has been shown that the widely used open-plan layout for offices, in which multiple workers must share a single open space, cannot effectively support their physical and psychological needs \cite{James2021}. Originally designed to maximise space efficiency and minimise desk-to-staff ratio as compared to traditional cellular layouts \cite{Kim2016}, open-plan layouts struggle to accommodate collocated work activities \cite{Jackson2009}, diverse personal needs \cite{Indraganti2021}, and varying situations and events \cite{Yildirim2007}. Consequently, open-plan layouts tend to introduce so-called \textit{environmental demands} to workers \cite{Awada2023}: visual \cite{Kim2013}, acoustic \cite{Haapakangas2018}, or privacy-related issues \cite{Lee2010} that `demand' workers to exert additional physical and/or psychological effort to achieve a desired work outcome \cite{Roskams2021}. Moreover, chronic exposure to these environmental demands increases the likelihood of physical and mental health deficits, burnout and low productivity \cite{Roskams2021}. It is not surprising then, how remote working from home gained immense popularity during the COVID-19 pandemic \cite{Levy2021}, despite its own set of `demands', such as the conflict between work and home responsibilities \cite{Bernhardt2022}, the extension of working hours that reduce physical activity \cite{Koohsari2021}, or the external perception of decreased productivity \cite{Galanti2021}. With knowledge-intensive companies now expecting employees to return to their offices to ensure longer-term team cohesion \cite{Souza2022} and maintain interest in collaboration \cite{Smite2021}, it has become crucial for open-plan layouts to support workers in coping with the environmental demands. This can be achieved by improving their \textit{environmental resources}, which are positive aspects in the workplace that workers can employ to reach their physical and mental needs \cite{Roskams2021}. It is the interplay of demands and resources that causes an architectural layout of an office to not only shape the socio-spatial behaviour of workers \cite{Lee2019, Lee2021, Alavi2019}, but also influence their physical and psychological health and wellbeing \cite{Nappo2019, Richardson2017}.


To tackle these aforementioned challenges, the so-called `flexible' office concept promises to allow the architectural layout to be customised according to a wide range of work activities \cite{Barath2022}. However, this flexibility still requires repetitive and cumbersome manual adjustments, which in turn cause social discomfort for workers \cite{Obrien2014} or require the active support from building managers who expect adjustments to be announced and planned beforehand. The `activity-based' office concept offers a collection of architecturally optimised spaces that are separately tailored to accommodate specific work activities \cite{Wohlers2017}. While an activity-based office circumvents many of the demands that are associated with open-plan office layouts, it however introduces different demands, such as the daily stress of finding a suitable workspace, the perception of not working efficiently when not sitting at a desk, and feeling detached due to the absence of personalised or private workstations \cite{Engelen2019}.  Given that each worker tends to prefer personal coping strategies to handle simultaneous demands and resources in their office \cite{Hong2020}, any architectural layout solution must also actively mediate between the personal needs and preferences of all workers who share the same workplace.


This research is built upon recent advancements in robotic furniture \cite{Ju2009, Sirkin2015}, which demonstrated the theoretical and technological feasibility of actively and even autonomously adapting an architectural layout via robotics. For example, previous studies have shown how robots (semi-)autonomously move everyday furniture elements like sofas \cite{Spadafora2016}, chairs \cite{Agnihotri2019}, footstools \cite{Sirkin2015}, trash cans \cite{Bu2023}, or spatial partitions \cite{Onishi2021, Nguyen2021} to enhance ergonomic \cite{Takashima2016} or privacy \cite{Onishi2022} demands, or even suggest new activities \cite{Agnihotri2019, Nguyen2020, Nguyen2022}. However, there exists little knowledge regarding the technical feasibility and longer-term usefulness of such robotic furniture in a real-world  environment. Moreover, it is fully unknown as to how robotic furniture should negotiate the different and often conflicting needs of people who must share the same environment. For example, in a typical open-plan office, workers tend to experience environmental demands that originate from their own co-workers, and experience these demands differently based on their personal sensitivities, ongoing work activity, or social relations with these co-workers. Contributing to prior HCI research addressing these challenges \cite{Fujita2023}, we conducted an in-the-wild study on how the control of a \textit{mobile robotic partition} is shared in an open-plan office. This partition is capable of actively adapting the office layout by autonomously manoeuvring between predetermined locations as selected by office workers, while avoiding both static and dynamic obstacles. We specifically chose an everyday-looking office partition, because its solid wall-like embodiment allowed it to change the visibility, connectivity, materiality and even the functionality of a space. Our research question is therefore: \textbf{How can the control of a mobile robotic partition be shared by multiple workers to mediate the environmental demands and resources of an open-plan office layout?}

Following a research-through-design approach \cite{Zimmerman2007}, we conducted a co-design session and a five-week in-the-wild study , during which we observed 13 workers as they utilised the mobile robotic partition. We recorded a total of 96 \textit{adaptations}, i.e. an instance where the robotic partition was deployed to adapt the spatial layout of the office, of which 77 were initiated by workers and 19 by the researchers using the Wizard-of-Oz method. Our observations revealed that workers appropriated four distinct spatiotemporal adaptation strategies that depend in terms of timing, i.e. whether the partition adapted a space to \textit{reactively} cope with an environmental demand, or to \textit{proactively} improve the environmental resources in preparation for a future demand; and location, i.e. whether the partition adapted close by the adaptation initiator or the demand.
By analysing the results from UEQ questionnaires \cite{Schankin2022}, digital logs, first-hand observations and in-depth interviews, we identified four types of environmental demands that the robotic partition was able to ameliorate: acoustic, visual, glare, and privacy; as well as six initiation regulating factors that influenced the selection of each adaptation strategy: the severity, provenance, and predictability of the environmental demand, and the potential, shareability, and intentionality of the adaptation.

This study thus offers three distinct contributions. Firstly, we developed a semi-autonomous mobile robotic partition able to navigate safely between predetermined locations in a real-world and therefore relatively unpredictable physical environment. Secondly, we deployed this robotic partition over a longer period of time to record how workers shared its control to ameliorate environmental demands and improve environmental resources. Lastly, we synthesised four spatiotemporal adaptation strategies, four design considerations and six initiation regulating factors that help suggest the most appropriate adaptation strategy.
By identifying when and how robotically-actuated adaptations should unfold in a real-world setting, these contributions aim to advance multiple smart building applications, which the potential of informing a future of semi-autonomous and even fully autonomous robotic furniture solutions that aim to adapt architectural layouts - according to the actual environmental demands and resources, as experienced by multiple collocated people.


\section{Related Work}
We base our study on prior knowledge of: 1) how workers in open-plan offices tend to cope with environmental demands or improve environmental resources; and 2) how people tend to perceive the intentions of robotic furniture.

\subsection{Coping Strategies in Open-Plan Offices}\label{sec:coping}
In a typical open-plan office setting, workers employ two primary coping strategies to deal with environmental demands: approach and avoidance \cite{Roth1986}. \textit{Approach coping} involves actively responding to, or seeking solutions for, demands; while \textit{avoidance coping} entails efforts to divert attention away from, or ignore, demands \cite{Skinner2003}. While avoidance coping may offer short-term relief, they are often linked to increased distress in the longer run, as unresolved environmental demands persist. On the other hand, approach coping that aims to neutralise the demand is more likely to lead to positive long-term psychological and physical health outcomes \cite{Taylor2007, Roth1986}.

A typical open-plan office provides various adjustable spatial elements that workers can utilise as environmental resources for approach coping, such as thermostats, windows, blinds, shades, or lighting \cite{Stazi2017}. However, these resources are not consistently used \cite{Obrien2014} because each worker usually avoids disrupting their ongoing work activity to deploy an adjustment, and only do so when the environmental demand exceeds their personal tolerance level \cite{Haigh1981}. When making such adjustments, workers often proactively aim to preempt potential future demands, like when they adjust the office lighting in anticipation of the daylight dimming \cite{Moore2003}, or when they close the blinds to avoid future distractions \cite{Bordass2001}. Faced with multiple environmental demands, workers prioritise the most pressing ones and sometimes resort to avoidance coping for less critical issues. For instance, they might endure visual glare to maintain a desirable view \cite{Sutter2006} or the exposure to natural daylight \cite{Day2012}. The frequency of using adaptive architectural elements is influenced by their accessibility \cite{Restrepo2020} and familiarity \cite{Indraganti2021}, with workers adjusting lighting more when its control is user-friendly \cite{Karjalainen2013} or conveniently located \cite{Lindelof2006, Bordass1995}. 

The decision for approach coping is influenced by the presence of others in the open-plan office \cite{Fanger1970, Galasiu2006}. Individual workers often feel less empowered to prioritise their needs, assuming someone else will handle the environmental demands \cite{Reinhart2003, Dorn2013}. The absence of personal adjustable spatial elements also hinders the adoption of approach coping \cite{Fabi2012, Karjalainen2013}, as adjustments that affect an entire office space are rarely used due to the perceived disruption to others \cite{Galasiu2006, Obrien2014}. In those cases, the location of the control plays an important role, as workers with better access are more likely to take action \cite{Day2012} by assuming the role of primary controllers \cite{Boyce2006}. Personal traits like personality or sociocultural background influence whether primary controllers share control or discuss changes with their colleagues \cite{Hong2020}. More strategic use of digital technologies stimulates approach coping, as LED lights that prompt individual workers to take action can be exploited as social validation to facilitate the increased use of windows \cite{Ackerly2012}, while gradually dimming lights can reduce conflicts among workers by minimising work disruption due to sudden lighting changes \cite{Lashina2019}.

\subsection{Initiating Adaptations via Robotic Furniture}\label{sec:roboticfurniture}
Expanding the concepts of `robotic building' \cite{Bier2014} and `architectural robotics' \cite{Green2016}, the field of robotic furniture \cite{Ju2009, Lello2011, Sirkin2015, Fink2014} introduces advanced robotic technology into everyday human environments \cite{Lello2011}. Recent advances demonstrate that robotic furniture is able to respond to the needs of people \cite{Jens2017}, and even can `nudge' their behaviour \cite{Fink2014} towards more ergonomic \cite{Fujita2021} or collaborative \cite{Takashima2015} benefits.

Robotic furniture was applied to reduce manual effort in rearranging spatial layouts \cite{Fallatah2021} through various control interfaces. Handheld controllers aided office workers in visualising immediate spatial changes caused by a robotic partition \cite{Onishi2022}, while voice commands benefited healthcare facility users with mobility constraints \cite{Brooks2021}. For users with no design background, selecting a predetermined location for robotic furniture to move to, from a preset list, was experienced similarly as manually indicating these locations via a click-and-drag gesture \cite{Stoddard2021}. When multiple people are involved, robotic furniture tends to utilise camera tracking techniques to identify and then optimise collective benefits, such as how a shape-shifting display partition adjusts its curvature based on the individual locations of people \cite{Takashima2016}, or a folding tabletop transitions between horizontal and vertical surfaces depending on the total amount of users \cite{Jens2017}.

\begin{figure*}[ht]
  \includegraphics[width=\linewidth]{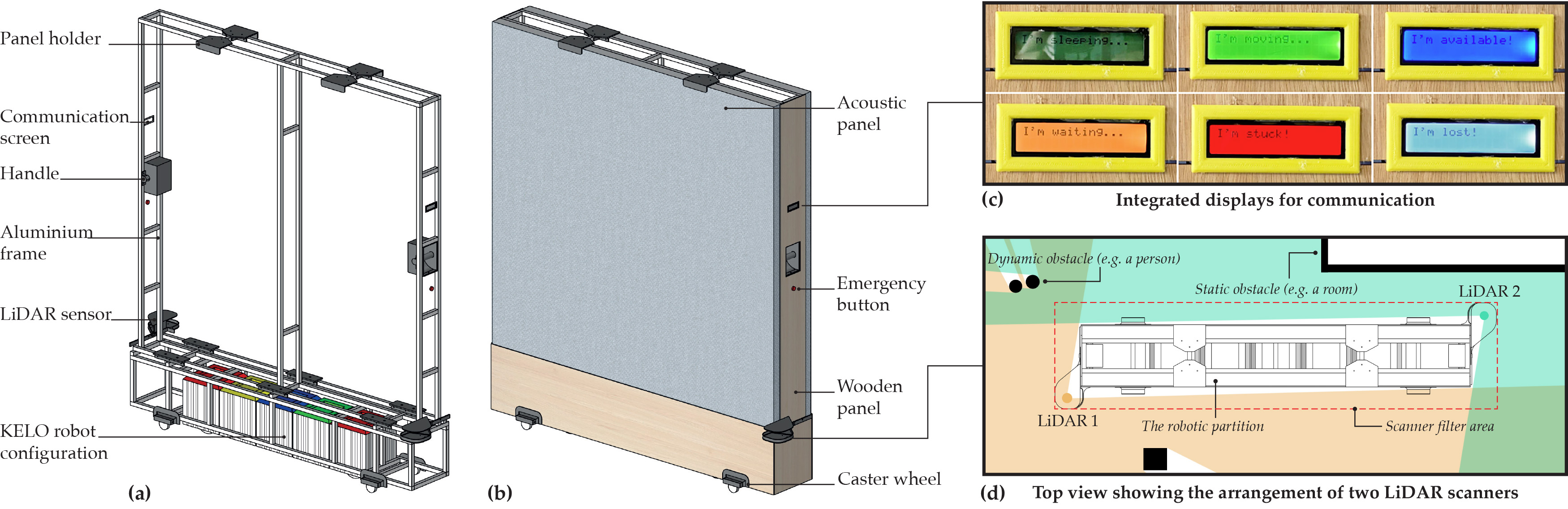}
  \caption{
  The robotic partition consisted of: 
  \textit{(a)} an (\(180x210x28cm\)) aluminium frame that housed a customised robotic configuration.
  \textit{(b)} The frame was covered by acoustic panels and wooden cladding.
  \textit{(c)} A small integrated LCD display at eye-height conveyed the internal system state.
  \textit{(d)} The arrangement of two \(LiDAR\) laser scanners enables a \(360\)-degree field of view to detect any obstacles.
  }
  \label{fig:technical}
  \Description{The technical implementation of the robotic partition. It consisted of: (a) an 180x210x28cm aluminium frame that housed a customised configuration of two robotic drive wheels, a battery with power distribution board, and a wireless control unit. Notably, two extrinsically-positioned LiDAR  laser scanners were required to ensure a 360-degree field-of-view around the wall's perimeter. (b) The frame was covered by two acoustic panels and wooden cladding, which included two handles for manual manoeuvring, two emergency stop buttons, and two small LCD displays. (c) The system state as conveyed by text and colour of the two small LCD displays, including 'sleeping', 'moving', 'available', 'waiting' corresponding to detecting a dynamic obstacle, and 'stuck' or 'lost' showing how the autonomous navigation was unable to reach its destination. (d) The arrangement of two LiDAR  laser scanners enables a merged 360-degree field of view to detect any obstacles, as the inaccessible view of each LiDAR  covered by the robotic partition can be captured by another.
  }
\end{figure*}

The dynamic movement of robotic furniture has been exploited to `nudge' \cite{Caraban2019} the behaviour of people. Anthropomorphic movements are able to attract human attention, like how a robotic chair invited passers-by to sit \cite{Agnihotri2019}, a robotic door encouraged passers-by to approach \cite{Ju2009}, or a toy-box robot prompted children to tidy their room \cite{Fink2014}. Well-considered non-anthropomorphic behaviour can support certain human activities without causing disruption, such as how a gradually inclining chair \cite{Fujita2021} or a slowly moving workstation \cite{Wu2018} discreetly encouraged healthier postures while working. However, people must be aware of these nudging capabilities to avoid being startled \cite{Gronvall2014}; and sufficient trust must have been built up by gradually proving the effectiveness of the robotic movements \cite{Lee2019Desk}. To support multiple collocated people, robotic furniture can move around their personal or shared social proxemic zones \cite{Krogh2017}, such as how a table subtly guided people to get closer for collaboration \cite{Takashima2015}, or how a spatial partition moved closer between people to influence their willingness to remain in its proximity \cite{Lee2013}.

\section{Methodology}
Our methodology includes the implementation of the robotic partition, the co-design session with the office workers, and the longer-term in-the-wild study. 

\subsection{Technical Implementation}\label{sec:technical}
The robotic furniture embodied a standard office partition with acoustic dampening features, and deliberately resembled commercially available options like Vitra's Dancing Wall \footnote{ Vitra Dancing Wall: \href{https://www.vitra.com/en-us/product/dancing-wall}{vitra.com}}, or Logovisual's Thinking Wall \footnote{ Logovisual Thinking Wall: \href{https://www.logovisual.com/product/whiteboards/thinkingwall/acoustics/thinkingwall-mobile-acoustic-freestander/}{logovisual.com}}. Its mobile functionality relied on the user-configurable industrial-level mobile service robot platform KELO, of which the drive wheels \footnote{ KELO Robile: \href{https://www.kelo-robotics.com/products/\#kelo-robile}{kelo-robotics.com}} are extremely silent, robust, and safe to use in human-occupied environments. 
To maintain a slim (\(28cm\) thick) profile, the two KELO drive wheels were positioned \(160cm\) apart on a single axis. This unique setup required several custom configurations, including: \textit{(1)} adding four caster wheels to prevent tipping; \textit{(2)} removing rigid connections between the aluminium frame and the robotic modules to prevent hyperstaticity issues \cite{Siegwart2011}; and \textit{(3)} adjusting the drive wheel firmware to accommodate their parallel orientation.

The semi-autonomous movement relied on six custom software modules atop 13 modules developed from existing libraries in the Robot Operating System (\(ROS\)) framework, such as \(cartographer\) for mapping, \(amcl\) for localisation, or \(teb\_local\_planner\) for navigation. Our developments incorporated several novel functionalities, such as merging two \(LiDAR\)  scanner data streams, interpreting the custom control interface input, recognising dynamic obstacles to halt movement and re-initiate navigation, and displaying the system status on two small LCD displays. The human safety measures restricted the speed to \(0.2\) to \(5cm/s\), which stopped for \(20\) seconds upon detecting any dynamic obstacle within a \(0.5m\) radius around any \(LiDAR\)  scanner. Two  clearly marked and easily accessible emergency buttons allowed the power of all the robotic modules to be switched off anytime.

\subsection{Architectural Context}

\begin{figure*}[ht]
  \includegraphics[width=\linewidth]{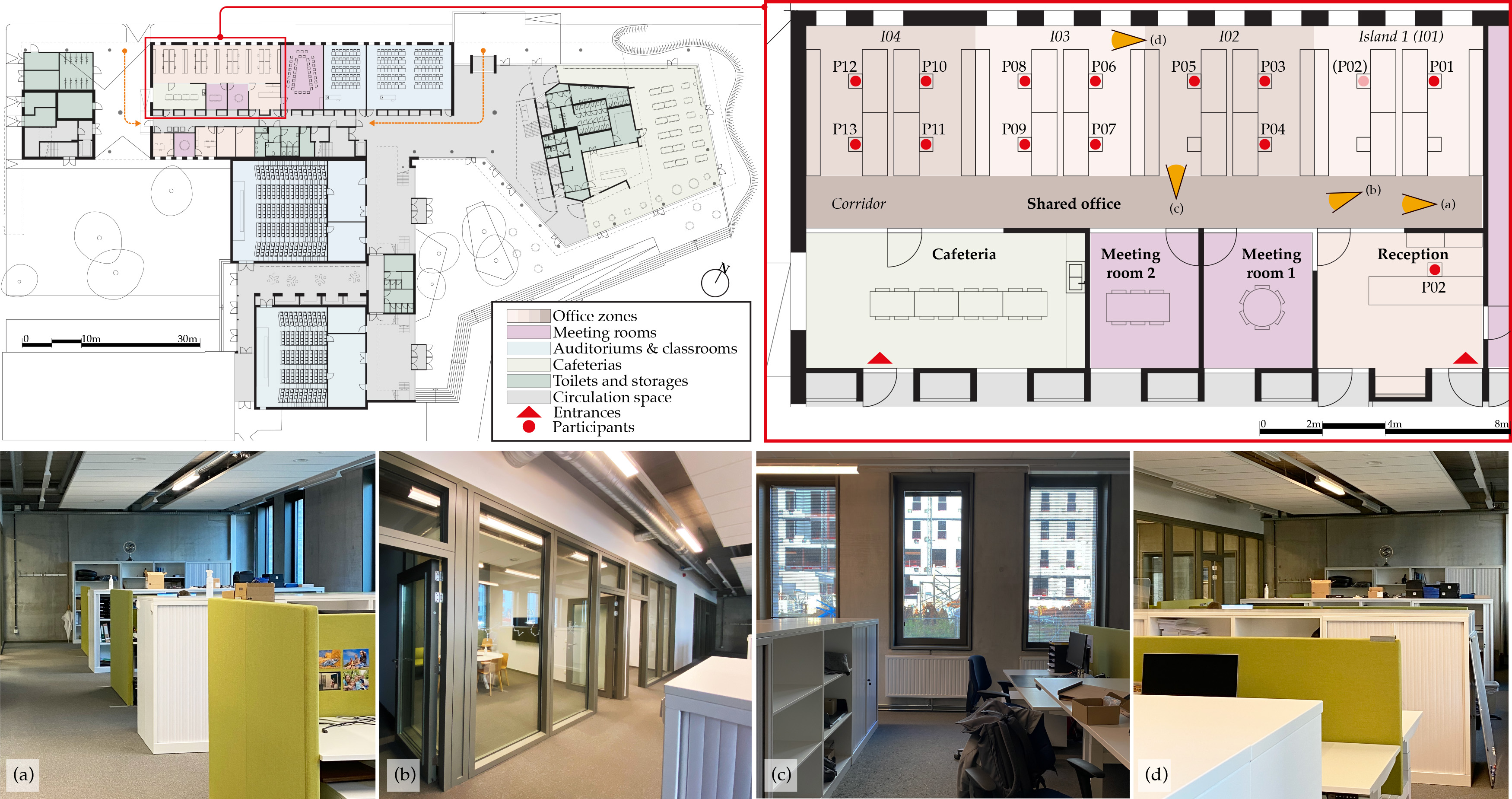}
  \caption{\textit{Top left:} The location of the open-plan office on the ground floor of a multi-functional university building. \textit{Top right:} The architectural layout of the open-plan office. \textit{Bottom:} Impressions of the architectural interior of the open-plan office.
  }
  \Description{Top left: The office environment is located at the North-West corner of the ground floor in a multi-functional university building that hosts an auditorium, a cafeteria, classrooms and office spaces. Top right: The office environment contains an open-plan office hosting four work islands, next to a common cafeteria, two meetings rooms and a semi-public reception area. Bottom: The open-plan office features a modest postmodern interior design that is characterised by a high ceiling, a sequence of narrow windows, and industrial materials with muted colours.}
  \label{fig:context}
\end{figure*}

Our study was conducted in an open-plan office located in a recently constructed, multi-functional building at our university campus. This building was chosen as our shared university affiliation significantly lessened the effort of obtaining ethical and operational permissions, as well as recruiting voluntary participants. We preferred this office because it housed the secretarial and administrative staff from a different faculty than our own, ensuring a more ecologically valid environment; and because its architectural layout provided sufficient space to accommodate the physical manoeuvring of the relatively large robotic partition.

As depicted in \autoref{fig:context}, the office environment comprises five distinct rooms: a large open-plan office space accommodating up to 16 people (\(120m^2\)), a semi-public reception area open to students (\(20m^2\)), a private cafeteria (\(36m^2\)), and two small private meeting rooms (\(12m^2\)) reservable by staff. While the open-plan office benefits from ample natural light through North-West oriented windows, the other four rooms only receive natural daylight through glass partitions with a direct view unto the open-plan office. To enable the \(LiDAR\)  scanners to detect these glass partitions, their lower part \(60cm\) had to be covered by paper sheeting, as illustrated in \autoref{fig:utilisation}.

The open-plan office hosts four distinct work islands (I01-04), each comprising two pairs of tables and a moderately high acoustic partition, and each separated by a set of large storage cupboards. These cupboards and partitions, both \(1.2m\) in height, obstruct direct views while sitting, yet still allow an overview of the entire space when standing or looking up. As the office was occupied by 13 workers (P01 to P13), including 12 women and one man, aged between 29 and 59 years (\(Mean=45.50,SD=10.48\)), the overall seating arrangement was determined by the type of work activities, as shown in \autoref{fig:context}. While I01 hosted an independent worker (P01) and the receptionist (P02); I02 accommodated workers (P03-P05) who often interacted with students; and I03 (P06-09) and I04 (P10-13) primarily engaged in focused work or remote meetings.

\subsection{Co-Design Session}
Thanks to the permission and organisation of the office manager, we first conducted an introductory presentation to familiarise all workers with our research goals, data gathering methods, and ethical precautions.
Subsequently, each office worker was invited to read and sign an ethical consent form that included the option to step out of the study at any time. 

\begin{figure*}[ht]
  \includegraphics[width=\linewidth]{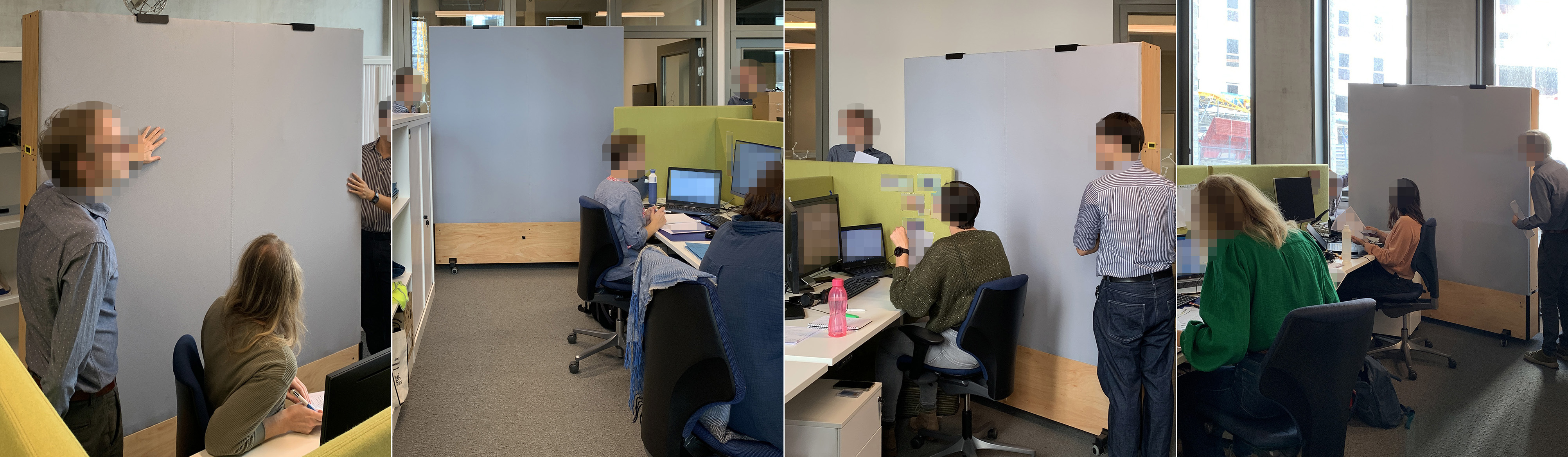}
  \caption{
  During the co-design session, 11 workers bodystormed with the researchers to determine how the robotic partition could potentially ameliorate specific environmental demands in their open-plan office.}
  \Description{Four photographs, showing from left to right how workers experienced first-hand the robotic partition, manoeuvred by two researchers, as it: provided shelter when standing in close proximity behind them; covered their view towards a distracting meeting room; improved privacy by creating a 'corner' at their workstation ; and reduced the glare from outside light on their computer display.}
  \label{fig:codesign}
\end{figure*}

\subsubsection{Procedure}
The 1.5-hour co-design session, facilitated by three researchers and attended by 11 participants (P08 and P10 were unable to attend), consisted of two phases.
During the 30-minute brainstorming phase, participants were invited to draw up to six hypothetical adaptations on a printed floor plan. A short questionnaire polled how each of their designed adaptations could potentially ameliorate an environmental demand. During the subsequent 60-minute bodystorming \cite{Oulasvirta2003} phase (see  \autoref{fig:codesign}), two researchers simulated each designed adaptation by manually manoeuvring the powered-down robotic partition, while discussing its expected feasibility and effectiveness with the participants. A third researcher observed and documented the think-aloud reflections \cite{Charters2003}.

\begin{figure*}[ht]
  \includegraphics[width=\linewidth]{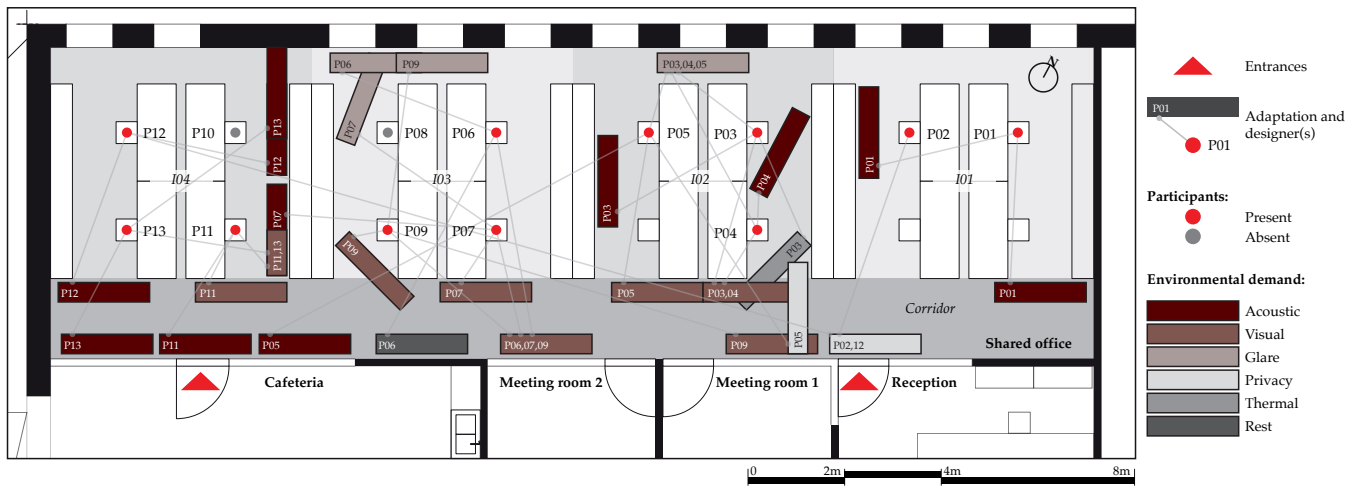}
  \caption{
  During the co-design session, 11 participants proposed 34 hypothetical adaptations, which were expected to ameliorate five different environmental demands. 
  }
  \Description{A floor plan with annotated locations of the robotic partition in 34 hypothetical adaptations, which were intended to cope with acoustic, visual, glare, privacy, and thermal demand. Each location is accompanied by an overlaid annotated network, illustrating its proxemic relation to the individual workstation of its author. Accordingly, participants in close proximity to the cafeteria (P10-13) and reception (P01) primarily addressed acoustic demands, while those near the corridor were more concerned about visual demands stemming from activities in meeting rooms, the cafeteria, or the reception. Participants near the windows (P03, P05, P06) experienced glare from sunlight, while privacy and thermal demands arose from unwanted visitor entry and airflow from the reception, respectively.
  }
  \label{fig:codesignresults}
\end{figure*}

\subsubsection{Preliminary Insights}
During the co-design session, the 11 participants proposed 34 (26 unique) adaptations to ameliorate five different environmental demands.
Participants expected the partition to tackle visual (\(n=12\), 35.3\%) and acoustic (\(n=11,32.4\%\)) demands caused by collocated work activities. 
They also positioned the partition to cover the sunlight glaring into the computer displays (\(n=6, 17.6\%\)), redirect cold ventilation flows between the entrances (\(n=1, 2.9\%\)), and enhance their privacy during breaks by preventing visitors to enter the office (\(n=3, 8.8\%\)). The participants also agreed on a visually and physically unobtrusive `resting' location when no adaptations were needed (\(n=1,2.9\%\)). 
\autoref{fig:codesignresults} highlights how the architectural context caused certain participants to prefer specific adaptations. For instance, participants located near the cafeteria (I04) or the reception (I01) mainly focused on acoustic demands, while those in the centre of the office (I02, I03) preferred to block visual demands from the meeting rooms, and those sitting next to the windows focused on coping with glare demands from the sunlight. While most participants focused on nearby demands, three participants (P05, P09, P12) focused on coping with demands that tend to happen further away from their workstations.

\subsection{In-the-wild Study}

\subsubsection{Adaptations}
\begin{figure*}[ht]
  \includegraphics[width=\linewidth]{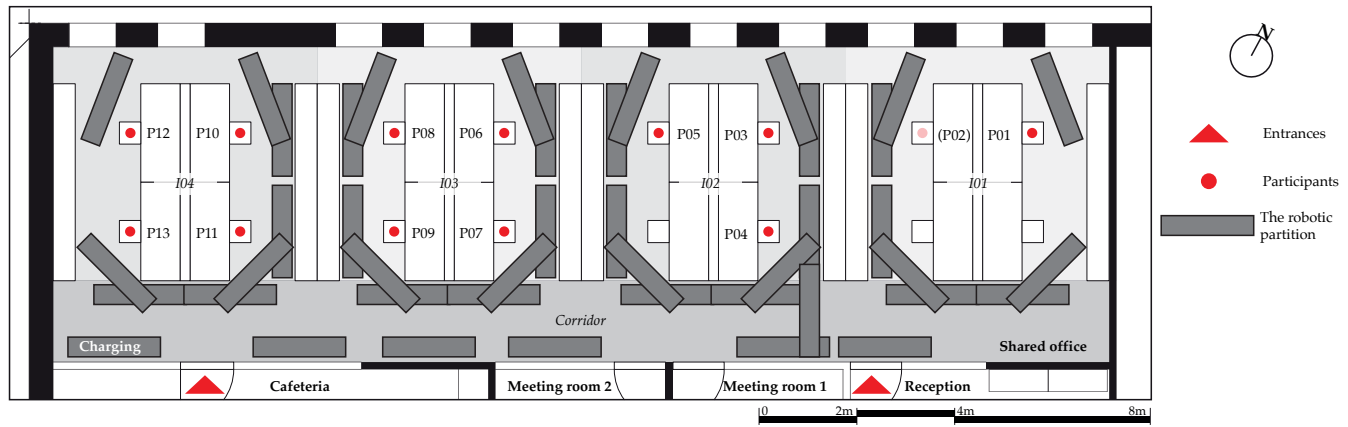}
  \caption{
  The 43 final adaptations that were integrated in the in-the-wild study were generalised from the 26 unique adaptations that were proposed during the co-design session.
  }
  \Description{A floor plan showing the locations of the robotic partition in the 43 final adaptations, generalised from the 26 proposed adaptations during the co-design session. Each work island is accompanied by up to 10 adaptations, with two to four dividing it from adjacent work islands, two covering its windows, and four sheltering it from the corridor. In addition, five adaptations covered the glass partitions next to the reception, the meeting rooms, and the cafeteria; while two adaptations covered the entrance from the reception to the open-plan office.
  }
  \label{fig:itwlocations}
\end{figure*}

\autoref{fig:itwlocations} shows the 43 final adaptations that were integrated in the in-the-wild study. While these final adaptations resulted from generalising the co-designed collection across the entire open-plan office, practical constraints led to excluding five of the 26 proposed adaptations: three adaptations that reduced glare by covering windows could not be reached by the robotic partition due to the narrow space between the desks and the outer wall; one adaptation (behind P10) was obstructed by electrical cables; while one adaptation covering the cafeteria door posed a fire escape hazard. The location for the `resting' location of the robotic partition was shifted closer to the cafeteria due to its accessibility to an electrical outlet for overnight charging.

\begin{figure}[ht]
  \includegraphics[width=\linewidth]{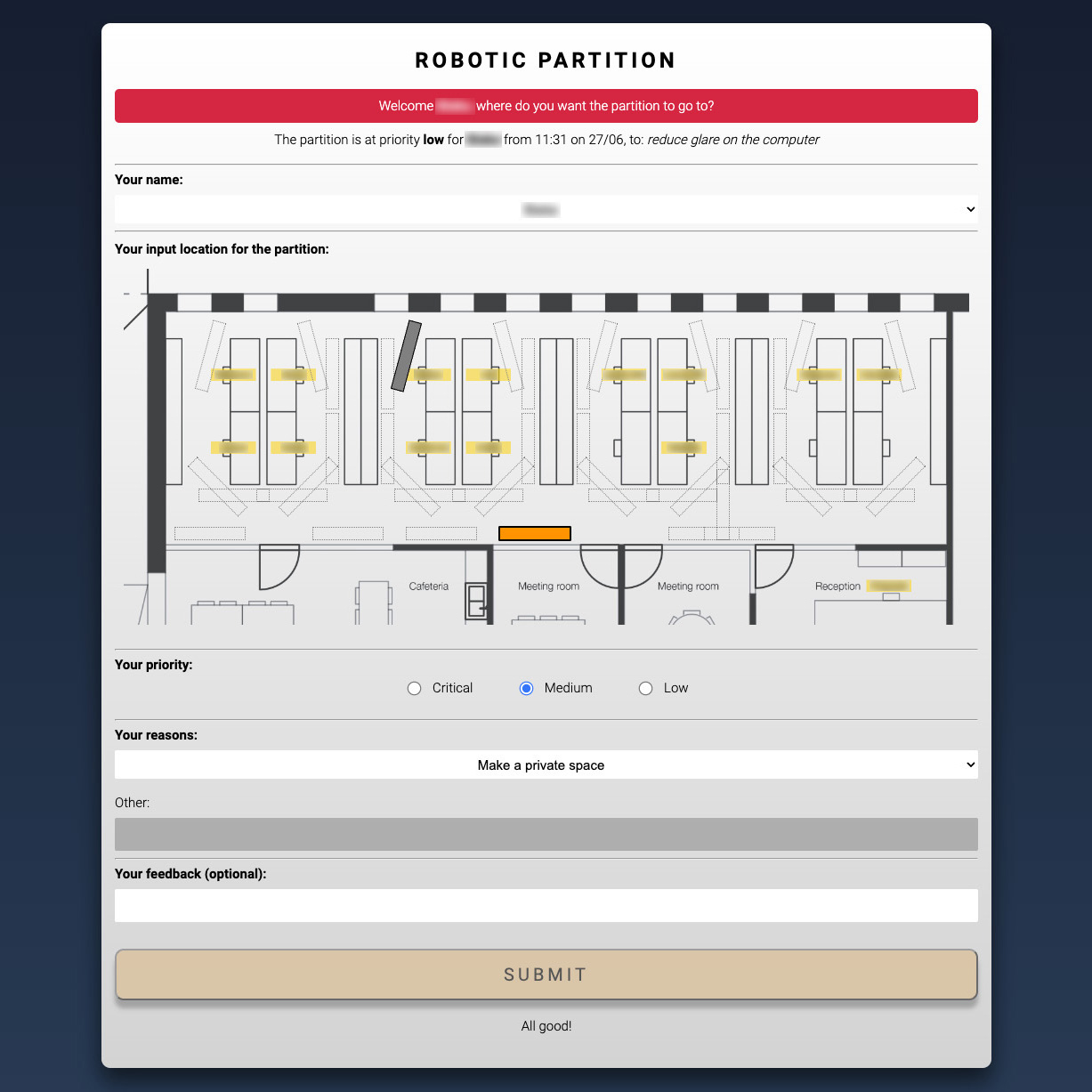}
  \caption{
  The custom web-based control interface allowed participants to initiate and comment on a new adaptation, or provide textual feedback on a previous adaptation. 
  }
  \Description{A screenshot of the customised web-based control interface. Participants could use this control interface to input their name (or act anonymously), select the appropriate adaptation location on the floor plan, choose its priority ranking, briefly describe the intention of the adaptation in a text field, and give their feedback on the previous adaptation to the researchers at any time. To limit the input effort, the intention description drop-down list was pre-populated with previously inputted intentions from that participant, while a new reason could be typed after selecting the option "Other".
}
  \label{fig:webpage}
\end{figure}

\subsubsection{Control Interface}
The custom web-based control interface (see \autoref{fig:webpage}) allowed participants to initiate an adaptation at any time. 
Recognising the social necessity for transparency when sharing the control of office adaptive technology \cite{Lashina2019}, the interface incorporated four key elements.
A \textit{status box} displayed the details of the active adaptation, such as the name of its initiator, timestamp, priority level (critical, medium, low), and intention. 
An \textit{input box} allowed any participant to leave their name, select a priority level, briefly outline their intention for starting a new adaptation, and select a new adaptation location from a floor plan featuring 43 rectangles representing possible adaptation locations. 
To limit the input effort, participants could select their name from a drop-down list which included an anonymous option, while the intention description provided a pre-populated drop-down list of their previous intentions.
Lastly, a \textit{feedback box} allowed participants to give feedback on whether their previous adaptations successfully met their intentions, which could be done at any moment without the need of a complete control request. 

\subsubsection{Procedure}
During a coffee break on the first day of the study, the primary researcher organised a 30-minutes workshop to demonstrate how to use the control interface, interpret the system status messages on the LCD display, and operate the emergency buttons. Participants were textually (through a three-page leaflet) and verbally instructed to contact the primary researcher in case the robotic partition displayed `lost' or `stuck' (see \nameref{sec:technical}), or showed any technical issue. They were also encouraged to give feedback on their initiated adaptations via the control interface as frequently as possible.

\begin{figure*}[ht]
  \includegraphics[width=\linewidth]{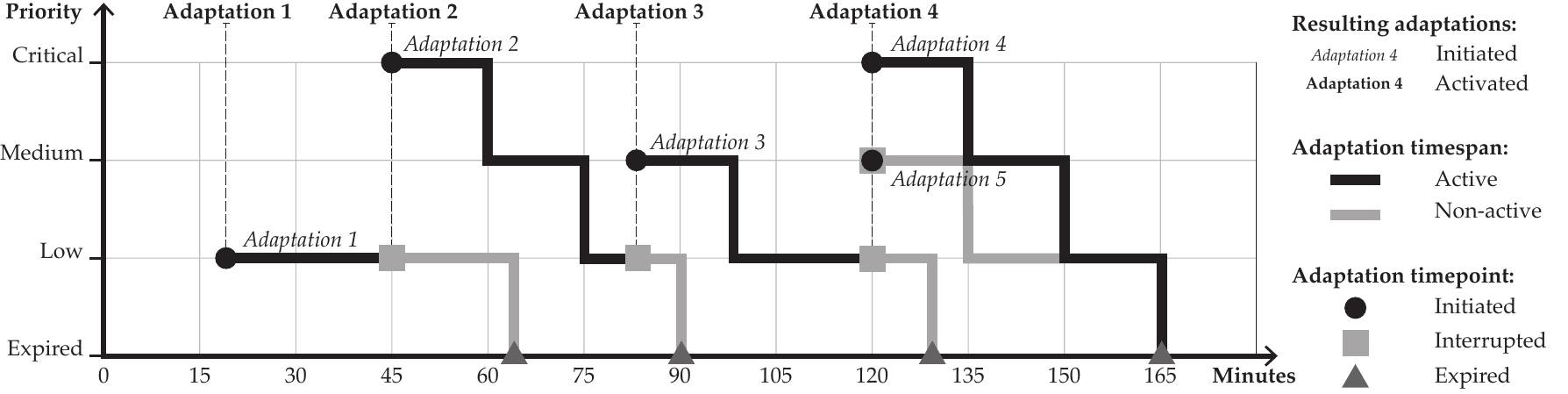}
  \caption{The priority levels that were submitted by participants were interpreted as \textit{interrupt priority levels} to determine the activation of the subsequent adaptation.}
  \Description{A diagrammatic example of how the priority levels were employed by the robotic partition software to determine the next adaptation. From left to right: an active adaptation with "low" priority was interrupted by a newly initiated adaptation with "critical" priority. After 30 minutes, the priority level of this adaptation was reduced to "low", and thus it was interrupted by another newly initiated one with "medium" priority. After 45 minutes, this adaptation expired. Among two later adaptations that were initiated at the same time, the one with "critical" priority was chosen to deploy instead of the "medium" one.
  }
  \label{fig:priority}
\end{figure*}

\begin{table*}[ht]
  \caption{
  Seven parameters extracted from the control log that characterised each adaptation.
  }
  \label{tab:parameters}
  \begin{tabular}{lllll}
    \toprule
    \textbf{Parameter} & \textbf{Description} & \textbf{Data type} & \textbf{Value / Unit} \\
    \midrule
    \textbf{Adaptation} & The selected adaptation of the robotic partition & Nominal & 43 adaptations \\
    \textbf{Initiator} & The participant initiating the adaptation (P01-13) & Nominal & 13 participants \\
    \textbf{Island} & The located work island of the initiating participant (I01-04) & Nominal & 4 work islands \\
    \textbf{Demand} & The environmental demand that the adaptation was intended to cope with & Nominal & 4 demands \\
    \midrule
    \textbf{Priority} & The priority level of the adaptation (low, medium, critical) & Ordinal & 3 levels \\
    \midrule
    \textbf{Time} & The timestamp of the adaptation initiation & Continuous & Hour \\
    \textbf{Duration} & The total time that the adaptation lasted & Continuous & Minute \\
    \bottomrule
  \end{tabular}
\end{table*}

The robotic partition control software determined whether to interrupt the currently active adaptation and/or activate a new one by interpreting the priority levels that were retrieved from the control interface as \textit{interrupt priority levels}. A subsequent adaptation was selected by its priority rank in the waiting list, which also included the active one. If multiple adaptations shared the same priority level, the one closest in physical distance to the current location of the robotic partition was preferred. To prevent any adaptation to dominate, the priority level of an active adaptation was automatically reduced every 15 minutes, as shown in \autoref{fig:priority}; and became `expired' after 45 minutes, after which the robotic partition remained at its location until the workday ended or a different adaptation was activated.


The primary researcher was physically present for at least one hour at the start (08:00) and end (18:00) of each workday during the deployment to check the office occupancy, make observations, turn on or charge the battery of the robotic partition. To simulate a fully autonomous behaviour, some adaptations were initiated remotely by the researchers.
These Wizard-of-Oz adaptations deliberately mirrored how participants had previously input their own adaptations before, and reflected the actual occupancy of the open-plan office. 


At the conclusion of the study, the primary researcher conducted (25 to 50 minutes) semi-structured interviews with each participant, resulting in over 7 hours of audio recordings. Before the interview, each participant completed a UEQ questionnaire \cite{Schankin2022}. The interview invited participants to reflect on how they experienced adaptations that were initiated by themselves (three questions), by colleagues (five questions), or via the Wizard-of-Oz method (five questions), how they utilised the control interface (three questions), and perceived the adaptations in their `static' state (five questions) versus when they were dynamically moving (six questions). 
Participants were also asked to assess the impact of each adaptation that they initiated, regarding whether it was ``\textit{effective}" or ``\textit{ineffective}" in addressing the intended environmental demand, by clarifying their input feedback (49 in 61 adaptations) and re-living their firsthand experience. 


\subsubsection{Data Acquisition and Analysis}

The study collected logs from the control interface, self-reported data from the interviews and questionnaires, and observational data such as notes and photographs by the primary researcher. Seven parameters extracted from the control logs (see \autoref{tab:parameters}) were triangulated with the qualitative analysis of the self-reported and observational data using the constructivist grounded theory methodology \cite{Charmaz2006}. 
The analysis process involved iterative open, focused, and theoretical coding. The open coding was conducted by the primary researcher, who considered the subjective reasoning and contextual nuances of the data collection that could only be interpreted through close observation. To mitigate biases, throughout the coding process, analytical memo writing and discussions among all authors were employed. The open coding initially generated 59 codes, which were refined into 27 categories representing abstract patterns among them during the focused coding. Subsequent group discussions and constant comparative analyses refined these categories, resulting in the six initiation regulating factors.



\section{Results}

\begin{figure*}[ht]
  \includegraphics[width=\linewidth]{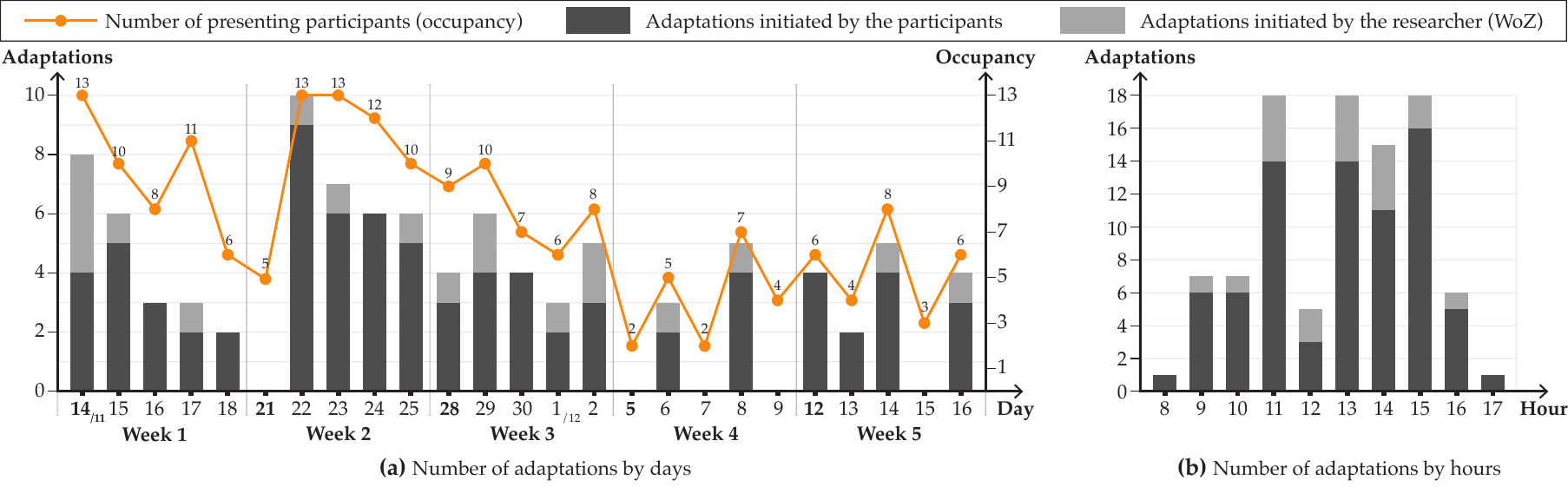}
  \caption{
  Most adaptations were initiated \textit{(a)} during the second week of the deployment (\(n=29\)) when participants became more familiar with the system (as explained by P01, P06, P11), and \textit{(b)} between the busiest working hours (11:00 to 15:00) to cope with with prevalent acoustic or visual demands, excluding the lunch break (12:00 to 13:00).
  }
  \Description{The timeline of initiated adaptations show that most adaptations occurred during the second week of the deployment (in total 29 adaptations) when most participants were present (10 to 13 participants a day) and had become sufficiently familiar with the system (explained by P01, P06, P11). In week 1 and 3, the number of adaptations was moderate (in total 22 adaptions each week, respectively), whereas week 4 and 5 saw a notable decrease in both the adaptation number (8 and 15 adaptations) and participant presenting (2 to 8 participants a day). Mapped among the working hours, the number of adaptations was the highest between the busiest hours (from 11:00 to 15:00, with a total up to 18 adaptations per hour), when acoustic or visual demands occurred the most, except during lunch time (12:00 to 13:00, 5 adaptations in total).
  }
  \label{fig:overview}
\end{figure*}

\begin{figure*}[ht!]
  \includegraphics[width=\linewidth]{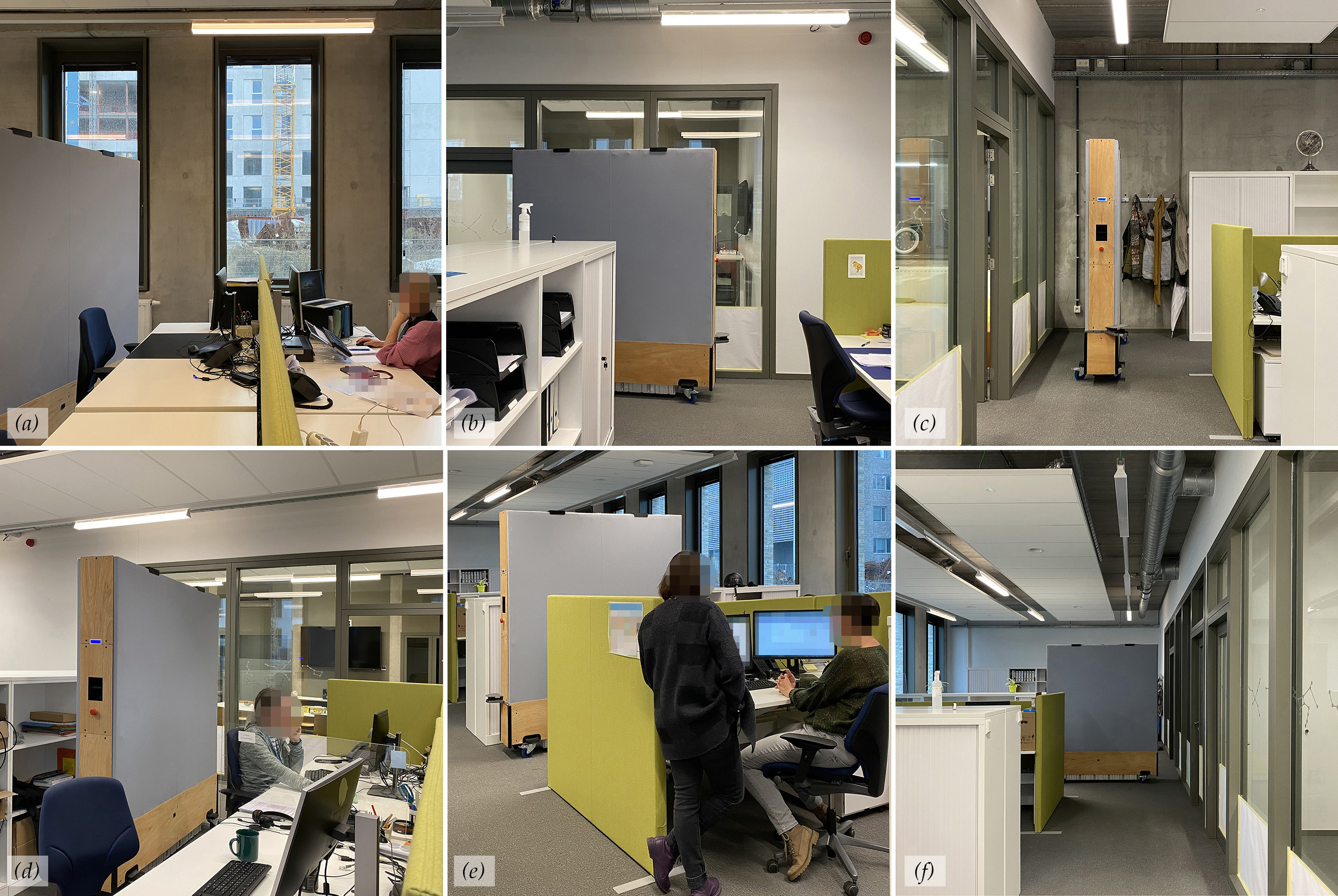}
  \caption{
  A collection of exemplary adaptations that were used to cope with the environmental demands in the open-plan office. 
  \textit{(a)} P01 covered the sunlight from the outside window to reduce glare. \textit{(b)} P04 covered the meeting room window before a meeting to enhance her privacy. \textit{(c)} P12 covered the cafeteria window to avoid observing an ongoing coffee break. \textit{(d)} P11 located the partition behind her to ``\textit{make a nice space}", which was the adaptation that lasted the longest (5 hours and 38 minutes). \textit{(e)} P13 covered the noise from a conversation between P07 and P03. \textit{(f)} P08 blocked the reception entrance to prevent visitors entering during a lunch break.
  }
  \Description{Six photographs showing example adaptations. (a) P01 covered the sunlight from the outside window opposite her workstation to reduce glare. (b) P04 covered the meeting room window before having a meeting inside to enhance her privacy. (c) P12 covered the cafeteria window on her right to avoid viewing a coffee break. (d) P11 located the partition right behind her to "make a nice space", which was the adaptation that lasted the longest (5 hours and 38 minutes). (e) P13 covered the noise from a conversation between P07 and P03 by locating the partition at their work island. (f) P08 blocked the reception entrance to prevent visitors entering during a lunch break.}
  \label{fig:utilisation}
\end{figure*}

As shown in \autoref{fig:overview}, a total of 96 adaptations occurred during the five-week in-the-wild study. This total count excludes the adaptations that were interrupted by their initiators within two minutes (\(n=12\)) or were automatically executed to move the partition to its `resting' location at the end of the day (\(n=31\)).
Participants initiated 77 adaptations (\(80.2\%\)), while the researchers initiated 19 WoZ adaptations (\(19.8\%\)), which taken together utilised only 27 of the 43 possible adaptation locations.
The adaptation durations varied widely (\(Mean=64.61, SD=60.93\) minutes), spanning from 5 minutes (P07) up to 5 hours and 38 minutes (P11, as seen in \autoref{fig:utilisation}(d)).

\begin{figure*}[ht]
  \includegraphics[width=\linewidth]{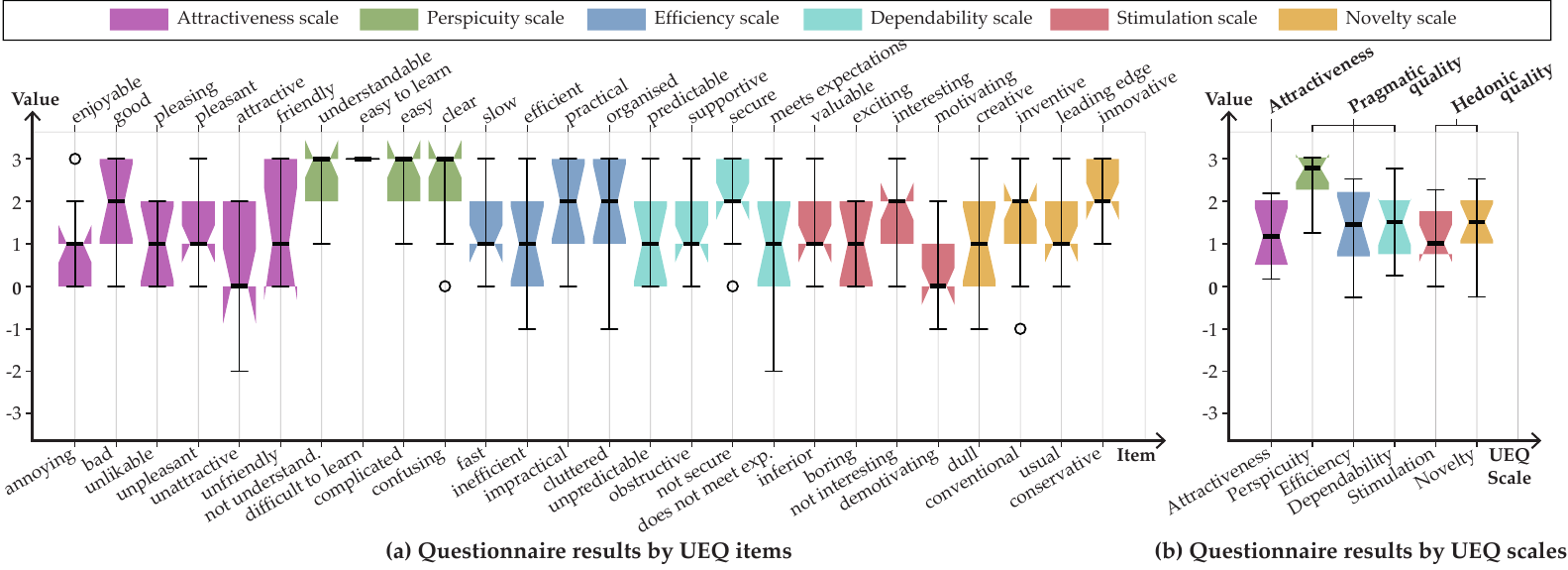}
  \caption{The UEQ questionnaire results. For readability reasons, each data item is formatted in a standardised way (i.e. negative terms at the bottom, positive terms at the top) instead of in the randomised order of questions that was presented to participants.}
  \Description{Box plots of the UEQ questionnaire results, documenting how participants ranked the robotic partition adaptations on attractiveness, pragmatic and hedonic qualities. For throughout explanation, see the Results section.}
  \label{fig:questionnaire}
\end{figure*}


\subsection{Questionnaire}\label{sec:questionnaire}
The UEQ results presented in \autoref{fig:questionnaire} indicate that participants ranked their experience with the adaptations ``\textit{above average}" across Attractiveness, Efficiency, Dependability, and Stimulation scales. Novelty received a rating of ``\textit{good}" (i.e. only 10\% of the UEQ benchmark dataset were better), whereas Perspicuity as ``\textit{excellent}" (i.e. within the top 10\% of the UEQ benchmark dataset).

Explaining their rankings, participants attributed Perspicuity (\(Mean=2.52,SD=0.36\)) and Efficiency (\(Mean=1.44,SD=0.82\)) to the design of the control interface (P01, P05, P09), LCD displays (P03, P04, P11), and the quick activation time (P03, P07, P13) after inputting a new adaptation.
Dependability (\(Mean=1.40,SD=0.84\)) was ranked based on the ``\textit{reliable}" (P11) movement of the partition: ``\textit{It moves slowly so you feel really safe around it, like it does not stop you from walking by}" (P12); its dynamic obstacle detection: ``\textit{It stopped immediately when it 'saw' me}" (P02); and robust navigation capability: ``\textit{It moves really well, even in small spaces}" (P09). 
Explaining Attractiveness (\(Mean=1.21,SD=0.62\), participants enjoyed that the robotic partition was ``\textit{helpful}" (P01) and did not give rise to additional environmental demands: ``\textit{It moved slowly, so I almost did not notice it at all while working}" (P03).
Regarding Novelty (\(Mean=1.46,SD=0.51\)), participants mostly found the partition ``\textit{innovative}" (P07) or ``\textit{professional}" (P02). 
In terms of Stimulation (\(Mean=1.12,SD=0.46\)), it was viewed as ``\textit{well-designed}" (P09) or ``\textit{worth an investment}" (P11); whereas some participants also disliked the robotic partition due to its plain appearance (P06), lack of autonomy (P10), or inflexibility (P08).



\begin{figure*}[p!]
  \includegraphics[width=0.95\linewidth]{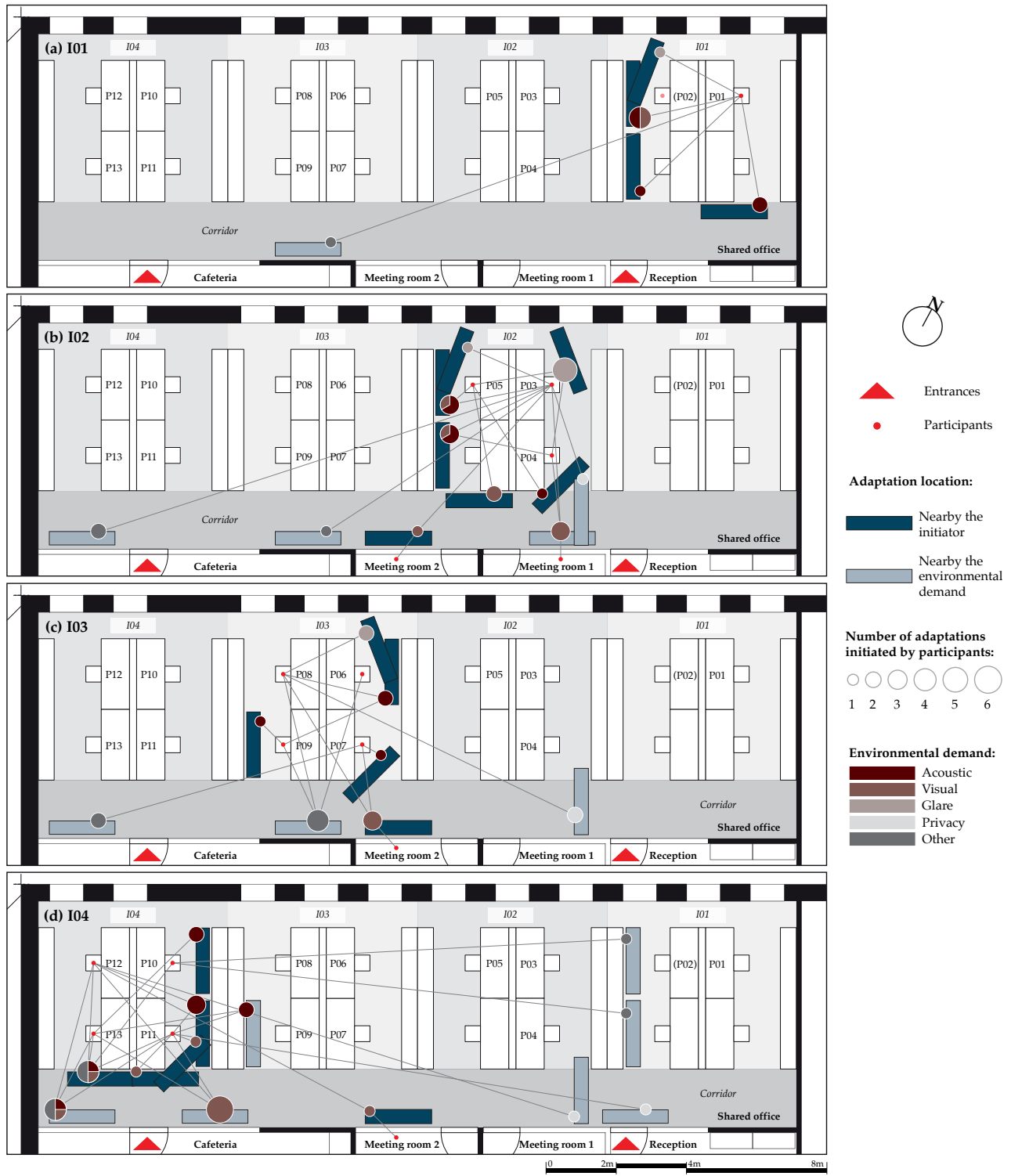}
  \caption{
  A map of all adaptation locations segregated by their  respective initiator 
  and classified by: (1) the environmental demand that they intended to cope with; and (2) their physical proximity to the location of the initiator versus the demand.
  }
  \Description{Four annotated floor plans showing the adaptations initiated by participants in each work islands. Accordingly, participants at work island I01 primarily coped with acoustic or visual demands by locating the adaptations around their work island; participants at work island I02 coped with visual demands from adjacent rooms by locating adaptations close to their work island and alongside the corridor; participants at work island I03 primarily coped with glare demands, but also initiated the most "resting" adaptations; and participants at work island I04 (P10, P11) felt empowered to locate some adaptations nearby other work islands (I03 and I01).
  }
  \label{fig:island1}
\end{figure*}

\begin{table*}[ht]
  \caption{
  How participants deemed the effectiveness of the robotic partition to cope with environmental demands. 
  }
  \label{tab:rirkfactors}
  \begin{tabular}{llcccc}
    \toprule
    \multicolumn{1}{c}{\textbf{Environmental}} & \multicolumn{1}{c}{\textbf{Adaptation as}} & \textbf{Initiated} & \textbf{Effective} & \textbf{Ineffective} & \textbf{Adaptations with}\\
    \multicolumn{1}{c}{\textbf{demand}} & \multicolumn{1}{c}{\textbf{environmental resource}} & \textbf{adaptations} & \textbf{adaptations} & \textbf{adaptation} & \textbf{unintended demand} \\
    \midrule
    Acoustic & Dampen sound from discussions & 23 & 15 & 4 & 5\\
    Visual & Cover view to activities & 23 & 19 & 2 & 5 \\
    Glare & Cover glare from the windows & 10 & 8 & 1 & 4 \\
    Privacy & Prevent visitors to enter & 5 & 5 && 2 \\
    \midrule
    & \multicolumn{1}{r}{\textbf{Total}} & 61 & 47 (77.0\%) & 7 (11.5\%) & 16 (26.2\%) \\
    \bottomrule
  \end{tabular}
\end{table*}

\begin{figure*}[ht]
  \includegraphics[width=0.6\linewidth]{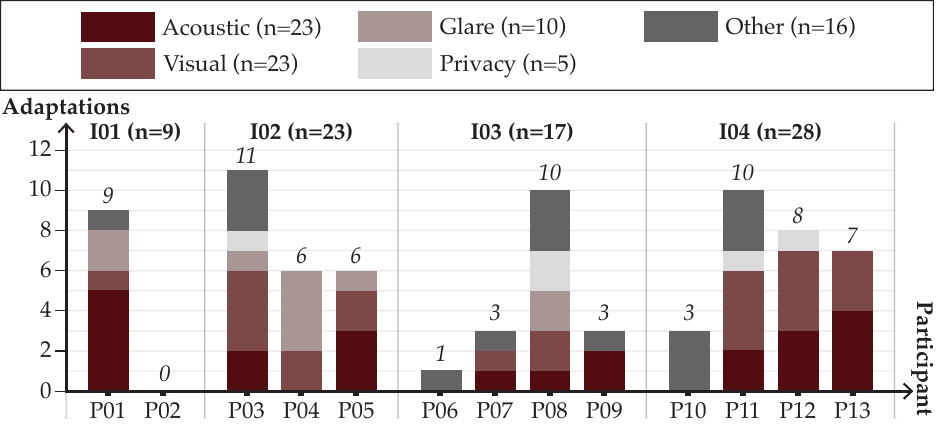}
  \caption{
  The 77 adaption initiations were unevenly distributed among the participants.
  Participants located near the entrance (I01) or the cafeteria (I04) initiated adaptations to cope with acoustic demands, whereas participants near the meeting rooms (I02) or the cafeteria (I04) focused more on visual demands.
  Three participants (P02, P06, P10) attributed their infrequent adaptation initiations to their natural ability to remain unaffected by environmental demands.
  }
  \Description{A bar chart showing the number of adaptations initiated by each participant to cope with acoustic, visual, glare, or privacy demands, or for other reasons. At I01, P01 initiated 9 adaptations (5 acoustic, 1 visual, 2 glare, and 1 other), whereas P02 did not initiate any adaptation. At I02, P03 initiated 11 adaptations (2 acoustic, 4 visual, 1 glare, 1 privacy, and 1 other), while P04 and P05 each initiated 6 adaptations (P04: 2 acoustic, 4 glare; P05: 3 acoustic, 2 visual, 1 glare). At I03, P06 initiated only 1 adaptation to return the robotic partition to 'resting' location, while P08 initiated 10 adaptations (1 acoustic, 2 visual, 2 glare, 2 privacy, and 3 other), and P07 and P09 each initiated 3 adaptations (P07: 1 acoustic, 1 visual, 1 other; P05: 2 acoustic, 1 other). Lastly, at I04, P10 initiated 3 adaptations for 'other' reasons, while P11 initiated 10 adaptations (2 acoustic, 4 visual, 1 privacy, and 3 other), P12 initiated 8 adaptations (3 acoustic, 4 visual, 1 privacy), and P13 7 adaptations (4 acoustic, 3 visual).
  }
  \label{fig:stressors}
\end{figure*}

\subsection{How Participants Assessed Adaptations}\label{sec:assessment}

As outlined in \autoref{tab:rirkfactors} and \autoref{fig:island1}, from the 61 adaptations (79.2\% of total 77 adaptations) that were initiated by participants to cope with an acoustic, visual, glare, or privacy demand, only 47 (77.0\% of 61) were deemed effective, whereas seven adaptations (11.5\%) were reported as ineffective. 16 adaptations (26.2\%), however, inadvertently introduced an unintended environmental demand even when they were effective in coping with the intended demand (\(n=9\)), such as by negatively affecting collaboration opportunities (\(n=5\)), available working areas (\(n=5\)), access to natural light (\(n=4\)), or circulation (\(n=2\)). Only one participant (P10) initiated two adaptations to cope with the unintended demand of feeling ``\textit{claustrophobic}", when his physical working area was reduced by previous adaptations initiated by P12 and P11, who had left the office.

14 adaptations (18.2\% of total 77 adaptations) were not meant to cope with an environmental demand. Accordingly, two adaptations tested how the robotic partition functioned to ``\textit{perceive how it feels like}" (P11), while 12 adaptations returned the robotic partition to its `resting' location to signal that it was available to use (P03, P07).


\subsubsection{Acoustic Demand}\label{sec:acoustics}
23 adaptations (37.7\% of 61) addressed acoustic demands that originated from the activities of other work islands (17 adaptations); from the cafeteria during coffee breaks (P11, P12, P13 - three adaptations); or from the reception when visitors were greeted (P01, P03, P05 - three adaptations).
Among these, eight participants appreciated the acoustic impact of 15 effective adaptations (65.2\% of 23): ``\textit{It really helped reduce the noise from behind me}" (P11), ``\textit{It covered the sound from the kitchen [cafeteria]}" (P12). However, four participants found four adaptations (17.5\%) ineffective in reducing noise due to the relatively limited height and width of the partition  (P02, P08, P10), or its inability to create a fully enclosed space (P04, P06).
Five adaptations (21.7\%) that were meant to address acoustic demands inadvertently disrupted collaborations (P01, P09, two adaptations), obstructed natural light (P07, two adaptations), or reduced the working area (P10, one adaptation), as reported by four participants.


\subsubsection{Visual Demand}\label{sec:visibility}
23 adaptations (37.7\%) addressed visual demands, mainly involving distracting activities like ``\textit{meetings}" (P04) or ``\textit{presentations}" (P03) in the meeting rooms (seven adaptations); ``\textit{coffee breaks}" (P11) in the cafeteria (seven adaptations); or circulation in the corridor (P05, P11, seven adaptations).
Among these, 11 participants considered 19 adaptations (82.6\%) effective, helping them ``\textit{focus more}" (P03) and ``\textit{work better}" (P11). However, two participants perceived the robotic partition as an unnecessary visual barrier (P06) since they believed the cupboards between the work islands (P10) served the same purpose.
Three participants (P03, P05, P09) highlighted how five adaptations (21.7\%) addressing the visual demands from the meeting rooms inadvertently introduced unintended demands for those inside the rooms by inadvertently disrupting their meeting (P05, P09) or reducing their access to natural light (P03).


\subsubsection{Glare Demand}\label{sec:glare}
10 adaptations (16.4\%) aimed to cover the glare caused by sunlight through the outside windows.
Five participants confirmed an effective reduction of glare in eight adaptations (80.0\%). However, one adaptation (10.0\%) failed to cover the light beams reflected by an opposite glass facade due to the limited height of the robotic partition (P01).
Due to their seating orientation in relation to sunlight direction, three participants (P04, P05, P08) occasionally placed adaptations at windows opposite their workstations. Four of these adaptations unintentionally created environmental demands for their colleagues by reducing ergonomic area (P04, P08 - three adaptations) or covering natural light (P05 - one adaptation). Consequently, two participants (P04, P08) refrained from initiating these adaptations when their opposite colleagues (P05, P06) were present.


\subsubsection{Privacy Demand}\label{sec:privacy}
Five adaptations (8.2\%) targeted privacy demands that arose from unwanted visitors during closing hours or lunch breaks, during which the robotic partition physically blocked the doorway between the reception and the open-plan office, as highlighted in \autoref{fig:utilisation}(f).
While all three initiators (P08, P11, P13) unanimously agreed that these adaptations (100.0\%) effectively enhanced their sense of privacy without disrupting the office circulation, two participants observed that two adaptations (40.0\%) inadvertently conveyed a ``\textit{wrong impression to the visitors}" (P13), signalling that ``\textit{they were not welcomed}" (P11).



\subsection{How Participants Initiated Adaptations}
We subsequently identified \textbf{six} regulating factors that empowered a typical participant to first decide when to initiate an adaptation (i.e. \textit{timing}), and then to determine its \textit{location} and \textit{priority level}. Decisions in terms of timing were categorised as either \textit{reactive}, i.e. initiated to cope with an ongoing environmental demand, or \textit{proactive}, i.e. initiated as an environmental resource to cope with a demand in the future.

\begin{table*}[ht]
  \caption{
  A quantitative analysis of the 61 participant-initiated adaptations suggests that the privacy demand was exclusively coped by proactive adaptations, whereas adaptations around the location of the initiator received a higher priority as well as took longer than the adaptations around the demand.
  }
  \label{tab:quantitative}
  \begin{tabular}{llr|rrrrr}
    \toprule
    \multicolumn{3}{c|}{\textit{\textit{Location}}} & \multicolumn{2}{c}{Initiator location} & \multicolumn{2}{c}{Demand location} \\
    \multicolumn{3}{c|}{\textit{\textit{Time}}} & \multicolumn{1}{c}{Reactive} & \multicolumn{1}{c}{Proactive} & \multicolumn{1}{c}{Reactive} & \multicolumn{1}{c}{Proactive} & \multirow{2}{*}{\textbf{SUM}} \\
    &&& \multicolumn{1}{c}{(\(n=32\))} & \multicolumn{1}{c}{(\(n=14\))} & \multicolumn{1}{c}{(\(n=9\))} & \multicolumn{1}{c}{(\(n=6\))} \\
    \midrule
    \textbf{Demand} && \textit{Acoustics} & 15 & 4 & 4 && \cellcolor{gray!25}23\\
    (adaptation && \textit{Visual} & 8 & 9 & 5 & 1 & \cellcolor{gray!25}23\\
    (num.) && \textit{Glare} & 7 & 3 & & & \cellcolor{gray!25}10\\
    && \textit{Privacy} & & & & 5 & \cellcolor{gray!25}5\\
    \midrule
    \textbf{Priority} && \textit{Low} & 19 & 9 & 5 & 4 & \cellcolor{gray!25}37\\
    (adaptation && \textit{Medium} & 7 & 1 & 4 & 1 & \cellcolor{gray!25}13\\
    num.) && \textit{Critical} & 6 & 4 && 1 & \cellcolor{gray!25}11\\
    \midrule
    \textbf{Duration} && \textit{Mean} & 80.7 & 72.4 & 58.5 & 39.7 & \cellcolor{gray!25}N/A\\
    (minutes) && \textit{SD} & 73.5 & 94.7 & 39.0 & 24.8 & \cellcolor{gray!25}N/A\\
    && \textit{Min} & 5.4 & 5.1 & 21.4 & 13.2 & \cellcolor{gray!25}N/A\\
    && \textit{Max} & 298.4 & 338.0 & 137.7 & 75.6 & \cellcolor{gray!25}N/A\\
    \bottomrule
  \end{tabular}
\end{table*}

\subsubsection{Demand Severity}\label{sec:severity}
Mentioned in 41 utterances by all 13 participants, the \textit{severity} of an environmental demand reflects how far it negatively impacted their work. Participants explained severity based on their personal sensitivities (n=3): "\textit{It is very difficult for me [...] to work in an open office, because I am easily distracted by sound}" (P11); proximity to the demand (n=4): "\textit{I'm usually distracted by people coming in from the entrance next to me}" (P01); or work requirements (n=6): "\textit{It was a very important meeting with a person from the government [...], so I did not want anyone to interrupt}" (P08).
Generally, participants preferred proactive adaptations to cope with future demands that can severely disrupt their work (n=6): "\textit{The most annoying thing is that you have to stop your focus to close the window [...] so I put the [partition] there beforehand}" (P05). They assigned higher priority levels to these proactive adaptations to prevent interruptions by others: "\textit{I made it critical so colleagues would not take the [partition] away during the [important] meeting}" (P08). 
However, when an unpredictable demand became too severe, they also felt empowered to initiate reactive adaptations (n=2): "\textit{I tried to ignore the distraction because I did not want to annoy the colleagues [...] but then they got too loud}" (P04).

\subsubsection{Adaptation Potential}\label{sec:potential}
Mentioned in 28 utterances by all 13 participants, the \textit{potential} of an adaptation describes how far they expected it to cope with an environmental demand. 
Four participants refrained from initiating adaptations because they predicted that the robotic partition might not adequately reduce noise (P06, P07), create a separate space (P10), or fully cover windows (P02). Three participants ceased initiating certain adaptations after firsthand experience revealed their ineffectiveness in addressing acoustic (P02, P08) or glare (P01) demands. 
Conversely, four participants initiated more adaptations after experiencing their effectiveness: "\textit{It felt very cosy sitting next to the [partition] [...], so I told it to move there again}" (P11); or through feedback from colleagues: "\textit{[P08] told me that the [partition] helped her with the reflection, so I tried it out and it worked!}" (P05).
Participants also compared environmental demand severity with adaptation potential to determine its timing and location. Potentially effective adaptations were proactively initiated around the location of the initiator to improve their environmental resources (n=6): "\textit{I made a quiet space for myself, just in case [...] it was very cosy because there was less distraction}" (P11); "\textit{I think a real wall should be installed [to cover the cafeteria] here because [the partition] helped me a lot to focus}" (P12). Meanwhile, adaptations that were potentially ineffective were rather initiated reactively at the location of the demand as a social signal (n=5): "\textit{[The partition] could not make a sound barrier for me, so I use it to send a message to colleagues to please help me focus}" (P13).

\begin{figure*}[ht]
  \includegraphics[width=\linewidth]{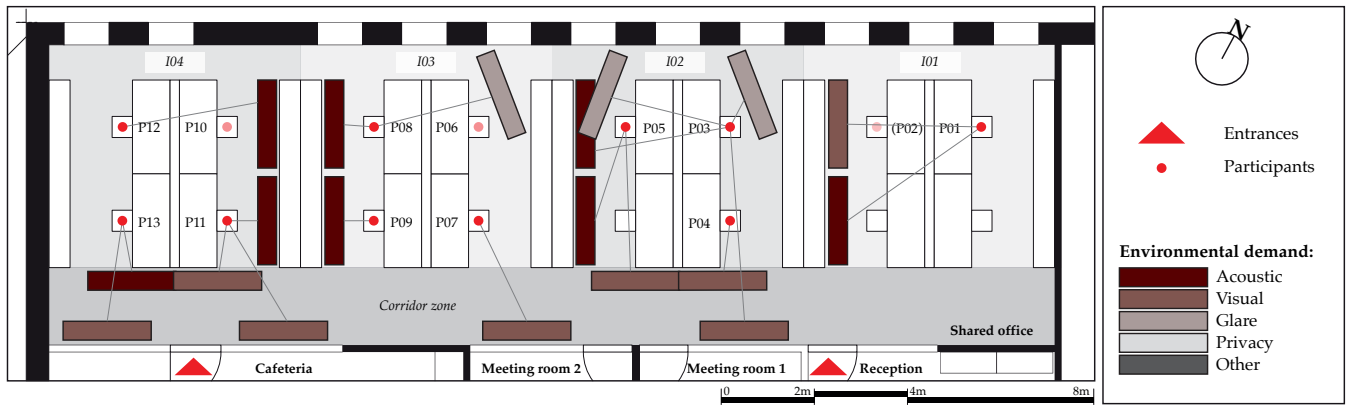}
  \caption{
  The 19 Wizard-of-Oz (WoZ) adaptations that were remotely initiated by the researchers appeared as being fully autonomous. The overlaid network foregrounds the participants that each WoZ adaptation aimed to support. 
  }
  \Description{An annotated floor plan showing the 19 WoZ adaptations, with the environmental demands they intended to cope with, and the participant they intended to help. Aligning with previous adaptations initiated by participants, adaptations coping with acoustic demands are mostly located between the islands. Meanwhile, adaptations coping with visual demands predominately covered the view to the corridor, the reception, the meeting rooms, or the cafeteria. Lastly, adaptations coping with glare demands were positioned to cover the windows.
  }
  \label{fig:wallinitiated}
\end{figure*}

\subsubsection{Demand Provenance}\label{sec:provenance}
The \textit{provenance} of an environmental demand, mentioned in 20 utterances by 11 participants, defines its cause: whether from external factors like glare and privacy demands, or from colleagues like acoustic and visual demands. 
To cope with demands originating from colleagues without causing social tension, participants preferred proactive over reactive adaptations (n=4): ``\textit{It is like I made the space for myself, instead of like [...] I had to use the [partition] because of you}" (P03); ``\textit{It was better to make my own corner beforehand so that I did not interrupt anybody's work}" (P12), choosing lower priority levels: ``\textit{You don't want to make it look so aggressive with critical priority}" (P13); ``\textit{I chose low priority to not offend my colleagues}" (P04), and positioning the adaptations around their immediate vicinity to improve their environmental resources: ``\textit{It is better to protect myself [...] I cannot expect everyone to stop their work to help me}" (P11).


\subsubsection{Adaptation Shareability}\label{sec:shareability}
Mentioned in 17 utterances by eight participants, the \textit{shareability} of an adaptation represents in how far its effectiveness could be shared with other colleagues.
While seven participants made this estimation on an individual level, e.g. ``\textit{I believed that we [I04] were all benefit from covering the noise}" (P12); one participant felt that everyone had to consent beforehand: ``\textit{Of course to stop the entrance of reception, everybody has to first agree}" (P08).
When perceiving that an adaptation would mutually benefit themselves and colleagues, participants felt empowered to initiate it reactively (n=3): `\textit{Their movements in the meeting room affected all three of us so I felt logical to use the [partition] at that moment}" (P05); in higher priority (n=3): ``\textit{I got the agreement from everyone, critical priority was logical for me}" (P08); towards the location of the demand (n=4): ``\textit{There was only one [partition] and it could not protect all three of us, so it was logical to move it to the other [work] island who was creating the distraction}" (P12); ``\textit{I could not find a good location at our [work] island for the [partition] to cover the view of both [P05] and me, so I chose to cover the meeting room instead}" (P09).

\subsubsection{Demand Predictability}\label{sec:predictability}
The \textit{predictability} of an environmental demand, mentioned in 16 utterances by six participants, reflects the likelihood of its occurrence. 
Participants anticipated glare demands based on weather conditions: ``\textit{It was a very sunny day [...] I get the [partition] first thing so the reflection will not disturb me}" (P04); acoustic or visual demands based on colleague presence: ``\textit{I made the [partition] stand there when it was almost full house [...] so that I could have my own space}" (P03); and privacy demands based on the academic calendar: ``\textit{It is a busy period so we expect some visitors at closing hours}" (P08). Based on these predictions, 11 participants initiated 20 proactive adaptations to improve their environmental resources in preparation for a future demand (see \autoref{tab:quantitative}).
However, the level of certainty in predicting a demand appeared to correlate to a sense of empowerment in determining whether to initiate a proactive adaptation: ``\textit{I was not sure how they were going to use the meeting room [...] and how it would disturb me [..] I chose not to cover it because it would be weird}" (P07), as well as its priority: ``\textit{I put the [partition] there just in case, so of course colleagues who need it more should take it over from me}" (P01); ``\textit{The distraction did not happen yet, so I used low priority [...]}" (P11).

\subsubsection{Adaptation Intentionality}\label{sec:intentionality}
Mentioned in 12 utterances by six participants, the \textit{intentionality} of an adaptation describes how readily other people can interpret its purpose. 
Participants interpreted the intention of an adaptation by considering its timing in relation to the environmental demand (n=6): ``\textit{She moved it when it was sunny so I knew it was the glare}" (P03), ``\textit{When it moved when somebody was talking, it was like a signal for them to keep their voice down}" (P08); as well as its location in relation to presenting colleagues: ``\textit{Because the [partition] is visible everywhere, it helps us to tell who needs support now}" (P01).
Consequently, participants considered reactive adaptations to more clearly convey their intentions (n=5): ``\textit{I did not want to put the [partition] behind her when I was not yet affected by sunlight}" (P08), ``\textit{Colleagues [in the cafeteria] understood that we moved it to protect us so they lowered their voice down}" (P11). 
To prevent potential misinterpretations of their intentions, participants avoided initiating adaptations close to their colleagues : ``\textit{I did not move [the partition] to other [work] islands [...] because it would be like I put my needs higher than others}" (P06).

\subsection{How Participants Experienced Wizard-of-Oz Adaptations}\label{sec:WoZ}

\begin{figure*}[ht]
  \includegraphics[width=0.75\linewidth]{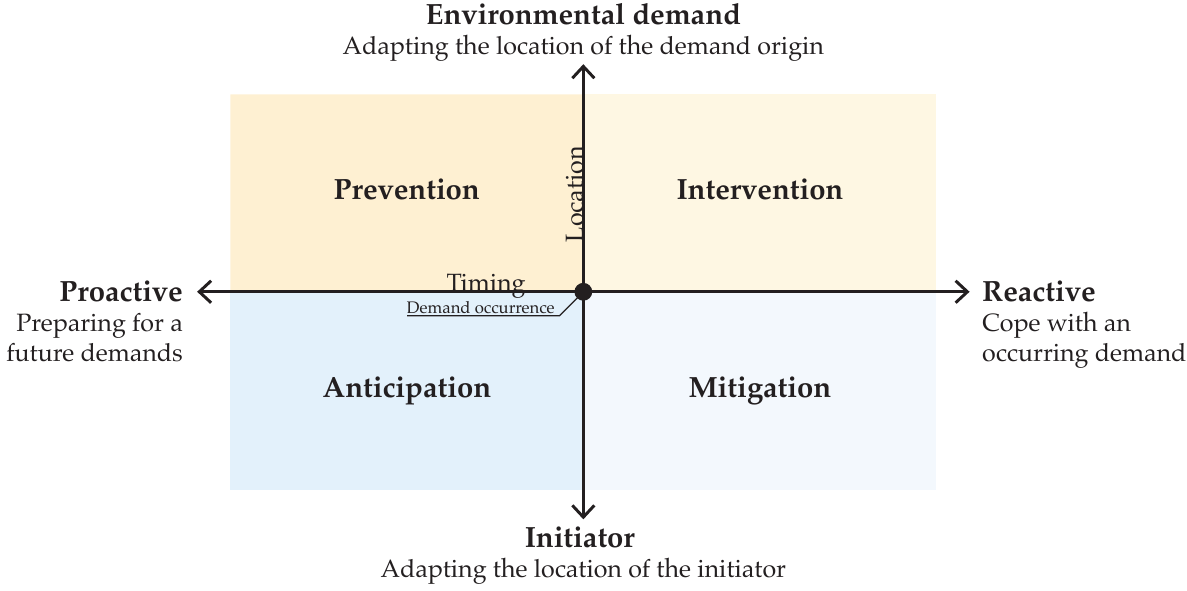}
  \caption{
  Participants initiated adaptations to cope with environmental demands via four distinct spatiotemporal adaptation strategies, which are determined by the \textit{timing} of the initiation and the \textit{location} of the adaptation, both  in relation to the environmental demand. 
  }
  \Description{A diagrammatic quadrant showing the four distinct spatiotemporal adaptation strategies. Accordingly, adaptations were either located nearby the environmental demand, or nearby the initiator who was suffering from it. Likewise, adaptations were either initiated before the demand had commenced, or while the demand was occurring. The four spatiotemporal adaptation strategies are therefore: prevention (proactive around the demand location), anticipation (proactive around the initiator location), intervention (reactive around the demand location), and mitigation (reactive around the initiator location).
  }
  \label{fig:strategies}
\end{figure*}

\autoref{fig:wallinitiated} illustrates the 19 Wizard-of-Oz (WoZ) adaptations that were covertly initiated by researchers.
Eight participants could not distinguish these adaptations, as they perceived all adaptations to address an environmental demand: ``\textit{It always went to a location that I think somebody would choose to help them at work}" (P08); and noted how the adaptations avoided creating unintended demands: ``\textit{I don't remember it ever moved to somewhere unreasonable}" (P01) and ``\textit{It would not go stand in the middle of a conversation}" (P04).
Three participants tolerated seemingly inappropriate WoZ adaptations as they assumed they were initiated by their colleagues: ``\textit{Sometimes it moved to strange locations, but I did not tell it to go away because I thought someone might need it there}" (P05); ``\textit{You saw where it was heading and you kind of assume that a colleague around there needed it to come}" (P07). This assumption was influenced by past experiences with adaptations bearing unintended demands: ``\textit{[P03] put the [partition] there before to help her focus, so I thought it was her}" (P05).
Only two participants (P06, P10) deduced that some WoZ adaptations were initiated by logical reasoning: ``\textit{All three of us were sitting in the cafeteria, so I assumed that you made [the partition] move}" (P06).


\section{Discussion}
We first synthesise all adaptations in four \textit{spatiotemporal adaptation strategies}, and then discuss how participants chose specific strategies by reasoning around the six initiation regulating factors.


\noindent From the analysis presented in \autoref{tab:quantitative}, we categorised the adaptations based on their \textbf{timing} and \textbf{location}. The initiation of an adaptation was timed either \textit{proactively} or \textit{reactively} in relation to the occurrence of the environmental demand. An adaptation was located either around the vicinity of the \textit{initiator} or the \textit{environmental demand}. Inspired by studies in medical science that modelled coping strategies of office workers to reduce stress \cite{Bishop2002, Subhani2018} or burnout \cite{McClafferty2014}, we similarly propose four spatiotemporal adaptation strategies as depicted in \autoref{fig:strategies}, including:

\begin{itemize}
    \item \textit{Prevention (n=6)}. Based on strategies aimed to prevent the occurrence of burnout before they occur \cite{McClafferty2014}, prevention involves initiating an adaptation proactively at the location of an environmental demand to reduce its likelihood of occurrence. For example, participants placed the robotic partition at an entrance to prevent any privacy demand from unwanted visitors (n=5).
    \item \textit{Anticipation (n=14)}. Inspired by strategies aimed at anticipating foreseeable stress \cite{McClafferty2014}, anticipation involves initiating an adaptation proactively around the location of the initiator, so that its environmental resource is sufficient improved to cope with a future demand. For instance, participants positioned the robotic partition adjacent to their workspace to create a cosy space that also covered potential visual demands (n=9).
    \item \textit{Intervention (n=9)}. Grounded on medical treatments aimed at enhancing immediate stress responses \cite{Bishop2002}, intervention involves initiating an adaptation reactively around the location of an ongoing demand, so to reduce its severity or even disrupt is occurrence. For example, participants placed the robotic partition around their colleagues who were making noise to convey their need for a quieter environment (n=4).
    \item \textit{Mitigation (n=32)}. Inspired by strategies aimed at immediately mitigating the health impact of stress \cite{Subhani2018}, mitigation involves initiating an adaptation around the initiator to improve its environmental resources in response to an ongoing demand. For example, participants deployed the robotic partition to cover the glare from a window next to them (n=7).
\end{itemize}

\noindent We discovered that the four spatiotemporal adaptation strategies varied sufficiently from one another, as participants carefully assessed six initiation regulating factors to determine the most appropriate adaptation strategy for a particular situation, as detailed in \autoref{tab:strategies}. Apart from considering timing and location, these strategies can be grouped by how prevention and intervention `attack' an ongoing or potential environmental demand, versus how anticipation and mitigation `protect' the initiator by improving the environmental resources around them.


\begin{table*}[ht]
  \caption{
  Given a situation, participants chose an appropriate spatiotemporal adaptation strategy based on their assessment of the six regulating factors, as condensed from our Results.
  }
  \label{tab:strategies}
  \begin{tabular}{lcccc}
    \toprule
    \multicolumn{1}{r}{\textit{Spatio (location)}} & \multicolumn{2}{c}{Initiator} & \multicolumn{2}{c}{Environmental demand} \\
    \multicolumn{1}{r}{\textit{Temporal (timing)}} & Proactive & Reactive & Proactive & Reactive \\
    \midrule
    \multicolumn{1}{r}{\textit{Strategy}} & \textbf{Anticipation} & \textbf{Mitigation} & \textbf{Prevention} & \textbf{Intervention} \\
    \midrule
    Demand severity & unbearable & bearable & unbearable \\
    Adaptation potential & effective & effective & ineffective & ineffective \\
    \midrule
    Demand provenance & colleagues & colleagues && external factors \\
    Adaptation shareability & individual && mutual & mutual\\
    \midrule
    Demand predictability & predictable & unpredictable & predictable & unpredictable \\
    Adaptation intentionality & & understandable & missunderstandable & missunderstandable \\
    \bottomrule
  \end{tabular}
\end{table*}

\subsection{How Spatial Adaptation was Appropriated as Social Interaction}\label{sec:socialinteraction}
Our results show that both the act of initiating and experiencing an adaptation were perceived as a form of social interaction, as participants selected an adaptation strategy by estimating how its spatial and temporal characteristics would be perceived by others.
As shown in \autoref{tab:strategies}, participants favoured adaptations around themselves (see \nameref{sec:provenance}) to avoid misrepresenting their intentions or causing disruptions among their colleagues (see \nameref{sec:intentionality}). However, when the impact of an adaptation could be collectively shared, participants felt empowered to place the adaptation around the demand as a deliberate act of assistance towards their colleagues (see \nameref{sec:shareability}). A reactive adaptation was preferred when its potential was estimated to be limited (see \nameref{sec:potential}), as reactiveness can be more easily interpreted by others, and thus become appropriated as a social signal towards colleagues who caused the demand (see \nameref{sec:intentionality}).

The strategic reasoning about location in our study resonates with insights from proxemics theories \cite{Hall1969, Krogh2017}, which evidenced that people perceive their occupied architectural space as their \textit{personal} proxemic zone \cite{Hall1969}. By initiating an adaptation around the location of a demand, an initiator extends their influence beyond their personal proxemic zone to directly impact others. Viewed through a Foucauldian lens \cite{Bradbury2008}, these adaptations represent a form of power negotiation, as an initiator seeks to shape socio-spatial dynamics, even aiming to influence the demand-causing behaviours of others. Given that people tend to avoid exerting power over the personal proxemic zones of others \cite{Krogh2017}, adaptation strategies breaching these zones were utilised with noticeably more caution to prevent social tensions. Our quantitative findings reflect this caution, with notably fewer instances of interventions and prevention (\(n=15\)) compared to mitigation and anticipation (\(n=46\)) adaptations. When participants initiated adaptations around their colleagues, often due to unbearable demand severity or mutual adaptation shareability (see \autoref{tab:strategies}), they opted for the intervention strategy, so to ease social tension as its immediate reactiveness is unlikely to become misinterpreted.

The caution of initiating adaptations might stem from the large dimensions and thus notable spatial impact of the robotic partition, or from the specific social dynamics within our participant cohort. However, we believe that our findings can extend to any human-building interaction technology intended to be shared by multiple people in a collocated setting, as the appropriation of any novel technology would similarly depend on not only its perceived effectiveness, but also its impact on colleagues and how it would be perceived as a social signal. Consequently, we propose that the relationship between the four adaptation strategies and the six regulating factors, as detailed in \autoref{tab:strategies}, can serve as a general guideline that informs how future human-building interaction can support individual workers to cope with environmental demands without disturbing their social community. 

\subsection{How Personal and Social Thresholds Inhibited the Initiation of Spatial Adaptations }\label{sec:threshold}
Our findings identified two inhibitors, termed as \textit{personal} and \textit{social thresholds}, which restrained participants from deploying spatial adaptation as \textit{approach coping} \cite{Skinner2003, Taylor2007, Roth1986}, and instead opting for \textit{avoidance coping} \cite{Skinner2003, Taylor2007, Roth1986} (see \nameref{sec:coping}).
The personal threshold emerged as participants deliberately ignored an environmental demand to prevent disrupting their ongoing work tasks, until the demand severity surpassed their prediction (see \nameref{sec:predictability}) or their coping ability (see \nameref{sec:severity}). The social threshold was evident as participants opted to adapt around their own location rather than around their colleagues to avoid social tension, as explained previously (see \nameref{sec:provenance}). These findings align with occupational health research in open-plan offices, indicating that workers address environmental demands with approach coping only when these demands surpass their ability to ignore them \cite{Leaman2001, Karjalainen2013}. Similarly, workers often ignore environmental demands to prevent disturbing colleagues or to allow others to take action \cite{Obrien2014, Dorn2013, Day2012}.

As approach coping causes a more positive impact on health and wellbeing compared to avoidance coping \cite{Raper2021, Taylor2007}, we propose that robotic furniture should focus on assisting approach coping by helping workers overcome their personal and social thresholds. 
Overcoming \textit{personal thresholds} could involve more proactive robotic behaviour, such as to suggest suitable adaptations to be initiated at strategic moments before the demand occurs, by way of subtle sounds or colour changes of the robotic partition \cite{Ackerly2012}.
Similarly, overcoming \textit{social thresholds} could involve communicating the intentions behind each adaptation to the people in its vicinity, such as through textual, visual or behavioural messages that could be intuitively interpreted.
We thus believe that these two thresholds can be transferred to support the future adoption of other human-building interaction technologies that need to engage with people close-by. Moreover, when these thresholds could already be addressed during the design phase, adaptations could take place as approach coping which has proven to bring more benefits.

\subsection{Contextual Interpretation of Spatial Adaptation}\label{sec:interpretation}
Our findings emphasise that participants primarily interpreted the intentions behind an adaptation based on its spatiotemporal context rather than the more explicit communication methods like the control interface, impromptu conversations, or corporate in-house messaging (see \nameref{sec:intentionality}). This contextual understanding mainly occurred by experiencing the adaptation first-hand, as participants compared its timing and location to the activities and locations of themselves and their colleagues (see \nameref{sec:WoZ}). However, because participants often assumed that the initiator benefited from their adaptation the most, they sometimes mistakenly interpreted adaptations that were meant to help colleagues (see \nameref{sec:shareability}).
This observation highlights a limitation in the identified four adaptation strategies, because they solely take into account the spatiotemporal relations between the initiator and the environmental demand, and overlook the fact that the initiator is a member of a community that tend to look out for one another (see \nameref{sec:intentionality}). This limitation might also be caused by the limited options in the control interface, which did not formally acknowledge the support towards others, and lacked a functionality to collectively initiate an adaptation.

As our Related Work section demonstrated (see \nameref{sec:roboticfurniture}), several robotic furniture studies already demonstrated how dynamic robot behaviour can be interpreted in a spatiotemporal way, even nudging specific human behaviour. 
Our study broadens these findings with an architectural perspective, demonstrating that a relatively large, solid and slow robotic partition is still able to convey its intentions not only by the timing of its `dynamic' movement but also by its relative `static' location to ongoing demands (see \nameref{sec:intentionality}). 
These findings thus highlight how adaptation, and most probably all other types of robots that might share human-inhabited spaces, should carefully consider not only how humans interpret the `dynamic' behaviour of robots through space, but also their provisional and temporal `static' postures between these movements.



\subsection{Towards Fully Autonomous Adaptation}\label{sec:autonomous}
While our study deployed a form of semi-autonomous robotic movement that still required direct human input to steer each location, we believe that our findings could provide insightful reflections to inform the design of fully autonomous adaptation systems in the future. For example, such autonomous systems could algorithmically or logically integrate the six initiation regulation factors to select the most appropriate adaptation strategy for a given situation (see \nameref{sec:socialinteraction}).
We thus envision that autonomous adaptation systems could potentially improve the health and wellbeing of workers by tailoring workplace environments much more towards coping with predictable as well as unpredictable environmental demands, while taking into account the needs and preferences of all individuals that are directly or indirectly affected.  
The same approach might also be useful to optimise the energy efficiency in unused parts of an open-plan office, or vice versa, `nudge' workers to utilise the available space more efficiently. However, as such a future system requires a continuous objective assessment of the six regulating factors that are fundamentally subjective, it still requires a human-in-the-loop feedback system to communicate, learn from, or correct its intentions.

It should be noted that steering the way humans interpret and behave in space has always been an inherent goal of architectural design. Architectural design has long leveraged how contextual meaning is communicated through design patterns \cite{Alexander1977} to influence how humans perceive space \cite{Lee2010} and place \cite{Harrison1996, Dourish2006}, utilise them \cite{Alavi2018}, or interact with others within them \cite{Lee2021}. However, architectural design uses contextual meaning to evoke spaces that are optimised towards specific activities, and ultimately counts on occupants to actively alter the spatial conditions when the activities change. Our approach of spatial adaptation yet draws upon an opposite premise, as we empowered architecture with a deliberate \textit{agency} to actively adapt the space around the occupants instead. Further research should demonstrate whether architectural design knowledge can facilitate more informed behaviour of robotically-actuated spatial adaptation, or whether there exists a still untapped design space of an ever-dynamic, always purposeful, and potentially even pleasurable robotic architecture that is as persistent and trustworthy as its conventional, yet static, counterpart.

\section{Limitations and Transferability}

Our \textit{participant cohort}, consisting mostly of women (12 of 13 participants) with an average age of 45 years, represents an underrepresented demographic in robotic research \cite{Hopko2022}. Their willingness to engage in an in-the-wild study, which involved documenting and observing their real-life, everyday work practice, provided us with valuable ecologically-valid insights into how everyday humans interact with a new type of robots. 
However, our participants also exhibited familiarity with a rather modernist view of architectural expression, a western-oriented office culture, and a normalised acceptance of automation that is typical in developed countries, all of which influence the perception of architectural space \cite{Mallgrave2015} as well as robots \cite{Esterwood2021}. Furthermore, the relatively harmonious social dynamics within our participant cohort might also have resulted in more considerate actions rather than conflicts, further highlighting the influence of workplace culture on accepting innovations or changes in their work environment.

Our \textit{qualitative-focused methodology} relied on self-reporting and self-reflecting by participants within their specific workplace, without cross-referencing with more objective sensor-based quantitative data, such as by objectively measuring the environmental demands. Furthermore, to maintain ecological validity, the researchers refrained from being extensively present, to ensure that participants would utilise the robotic partition naturally. As cameras and wearables were not deployed to respect the privacy of the real-world work environment, this study lacks more fine-grained behavioural observations. Given that our research question focuses on capturing personal and firsthand experiences, we provided a thick and rigorous description of our results so to maximise its transferability. 

The \textit{physical embodiment} of the robotic partition prototype, with its considerable physical size and solidity aimed to resemble an impactful architectural element, ultimately limited its navigation capabilities.
While the solidity was crucial for evoking feelings of shelter and protection (see \nameref{sec:potential}) in comparison to lightweight curtains or whiteboards; its mobility was integral to our research question into how a truly movable architectural element could be used much more effortlessly. Notably, seven participants expressed how the robotic partition was able to improve their workspace experience compared to their conventional practice of moving to the meeting room to cope with a disturbing environmental demand. They found that our robotic partition allowed them to work comfortably at their own workstations without relocating their equipment (P03, P11, P05, P13), facilitated team cohesion by assisting others with adaptations (P04, P08, P12), and alleviated feelings of isolation within separate spaces due to individual sensitivities (P11). 

Finally, our study, conducted through an exploratory research-through-design approach, recognises that the use of our prototypical robotic partition may not represent the optimal solution for addressing environmental demands in this particular open-plan office. While lighter, more flexible robotic partitions \cite{Onishi2022, Lee2013} could potentially better address visual or glare demands, other existing adaptive technologies such as smart lighting \cite{Lashina2019}, smart ventilation systems \cite{Seitz2017}, or kinetic acoustic ceilings \cite{Geoffrey2012} may effectively tackle different environmental demands that a robotic partition might not be equipped to handle. However, as our study highlights the shared nature of how workers typically adopt adaptive solutions, we propose that future research should determine the scope of our findings to these alternative technologies.

\section{Conclusion}
In this study, we investigated how multiple workers adapted their open-plan office layout by sharing the control of a mobile robotic partition that semi-autonomously moved between predetermined locations to cope with acoustic, visual, glare, and privacy demands. We observed how they deployed four distinct spatiotemporal adaptation strategies that differ in terms of timing, i.e. whether the partition adapts a space before or during the occurrence of an environmental demand; and location, i.e. whether the partition adapts around the proximity of the initiator or the demand itself. We identified six initiation regulating factors that determined the selection of an appropriate spatiotemporal adaptation strategy, including the severity, provenance, and predictability of the environmental demand, and the predicted potential, shareability, and intentionality of the adaptation. We discussed how participants appropriated spatial adaptation as a form of social interaction, while its utilisation as approach coping with environmental demands was inhibited by their personal and social thresholds. We reflected on the capability of spatial adaptation to convey intentionality through the interpretation of its surrounding spatiotemporal context, and briefly imagined the applicability of our findings towards the development of future fully autonomous adaptation systems.
With these insights, we aim to inspire human-building interaction and robotic furniture to address the inherent yet often overlooked challenge of assisting individual occupants in coping with environmental demands in situations where other occupants may either be the source of the demand or be impacted by the chosen coping strategy. Addressing this challenge has the potential to yield positive outcomes, not only in terms of comfort and productivity, but also in augmenting the longer-term health and wellbeing for occupants.

\begin{acks}
We would like to thanks our participants for accommodating our research to occur at their office, as well as their time, effort, and valuable insights. This research is supported by the KU Leuven ID-N project IDN/22/003 ``Adaptive Architecture: the Robotic Orchestration of a Healthy Workplace".
\end{acks}

\bibliographystyle{ACM-Reference-Format}
\balance
\bibliography{sample-base}

\end{document}